\documentclass[twocolumn,twocolappendix]{aastex63}

\usepackage{graphicx}	
\usepackage{xspace}
\usepackage{savesym}
\savesymbol{tablenum}
\usepackage{siunitx}
\restoresymbol{SIX}{tablenum}
\usepackage{booktabs}
\usepackage{placeins}
\usepackage{float}
\usepackage{amsmath}
\usepackage{tabularx}

\newcommand{\spitzer}{\textit{Spitzer}\xspace}
\newcommand{\msun}{\hbox{M$_{\odot}$}}
\newcommand{\herschel}{\textit{Herschel}\xspace}
\newcommand{\gaia}{\textit{Gaia}\xspace}

\newcommand{\kms}{\rm km\,s^{-1}}
\newcommand{\pc}{{\rm pc}}

\newcommand{\planck}{\textit{Planck}\xspace}

\newcommand{\mum}{\micron\xspace}
\newcommand{\degree}{\mbox{$^{\circ}$}\xspace}
\newcommand{\nh}{$N_{\rm H}$\xspace}
\newcommand{\td}{T_{\rm d}\xspace}

\newcommand{\nhp}{N$_2$H$^+$\xspace}
\newcommand{\co}{C$^{18}$O\xspace} 
\newcommand{\hco}{HCO$^+$\xspace}  
\newcommand{\sco}{$^{13}$CO\xspace}

\received{October 17, 2020}
\revised{December 1, 2020}
\accepted{December 15, 2020}
\submitjournal{ApJ}

\shorttitle{California filament rotation}
\shortauthors{Álvarez-Gutiérrez et al.}
\graphicspath{{./}}

\begin{document}

\title{Filament rotation in the California L1482 cloud}

\correspondingauthor{Rodrigo H. Álvarez-Gutiérrez}
\email{rodralvarez@udec.cl}

\author{R. H. Álvarez-Gutiérrez}
\affiliation{Departamento de Astronom\'{i}a, Universidad de Concepci\'{o}n, Casilla 160-C, 4030000 Concepci\'{o}n, Chile}

\author{A. M.\ Stutz}
\affiliation{Departamento de Astronom\'{i}a, Universidad de Concepci\'{o}n, Casilla 160-C, 4030000 Concepci\'{o}n, Chile}
\affiliation{Max-Planck-Institute for Astronomy, K\"{o}nigstuhl 17, 69117 Heidelberg, Germany}

\author{C.Y. Law}   
\affiliation{Department of Physics, The Chinese University of Hong Kong, Ma Liu Shui, Shatin, NT, 852 Hong Kong SAR}
\affiliation{Department of Space, Earth \& Environment, Chalmers
  University of Technology, SE-412 96 Gothenburg, Sweden}

\author{S. Reissl}         
\affiliation{Universit\"{a}t Heidelberg, Zentrum f\"{u}r Astronomie, Institut f\"{u}r Theoretische Astrophysik, Albert-Ueberle-Str. 2, 69120 Heidelberg, Germany}

\author{R. S. Klessen}  
\affiliation{Universit\"{a}t Heidelberg, Zentrum f\"{u}r Astronomie, Institut f\"{u}r Theoretische Astrophysik, Albert-Ueberle-Str. 2, 69120 Heidelberg, Germany}
\affiliation{Universit\"{a}t Heidelberg, Interdisziplin\"{a}res Zentrum f\"{u}r Wissenschaftliches Rechnen, Im Neuenheimer Feld 205, 69120 Heidelberg, Germany}

\author{N. W. C. Leigh}
\affiliation{Departamento de Astronom\'{i}a, Universidad de Concepci\'{o}n, Casilla 160-C, 4030000 Concepci\'{o}n, Chile}
\affiliation{Department of Astrophysics, American Museum of Natural History, New York, NY 10024, USA }

\author{H.-L. Liu}       
\affiliation{Departamento de Astronom\'{i}a, Universidad de Concepci\'{o}n, Casilla 160-C, 4030000 Concepci\'{o}n, Chile}
\affiliation{Department of Physics, The Chinese University of Hong Kong, Ma Liu Shui, Shatin, NT, 852 Hong Kong SAR}
\affiliation{Chinese Academy of Sciences South America Center for Astronomy, China-Chile Joint Center for Astronomy, Camino El Observatorio \#1515, Las Condes, 7591245 Santiago, Chile}

\author{R. A. Reeves}
\affiliation{Departamento de Astronom\'{i}a, Universidad de Concepci\'{o}n, Casilla 160-C, 4030000 Concepci\'{o}n, Chile}

\begin{abstract}
We analyze the gas mass distribution, the gas kinematics, and the
young stellar objects (YSOs) of the California Molecular Cloud (CMC)
L1482 filament. The mean \gaia DR2 YSO distance is
511$^{+17}_{-16}$~pc.  In terms of the gas, the line-mass (M/L)
profiles are symmetric scale-free power-laws consistent with
cylindrical geometry.  We calculate the gravitational potential and
field profiles based on these. Our IRAM~30~m multi-tracer
position-velocity diagrams highlight twisting and turning
structures. We measure the \co velocity profile perpendicular to the
southern filament ridgeline. The profile is regular, confined
(projected $r\lesssim0.4$~\pc), anti-symmetric, and to first order
linear, with a break at $r\sim0.25$~\pc.  We use a simple solid-body
rotation toy model to interpret it.  We show that the centripetal
force, compared to gravity, increases toward the break; when the ratio
of forces approaches unity, the profile turns over, just before the
implied filament breakup.  The timescales of the inner (outer)
gradients are $\sim\,$0.7 (6.0)~Myr.  The timescales and relative
roles of gravity to rotation indicate that the structure is stable,
long lived ($\sim$ a few times 6~Myr), and undergoing outside-in
evolution. This filament has practically no star formation, a
perpendicular \planck plane-of-the-sky (POS) magnetic field
morphology, and 2D ``zig-zag'' morphology, which together with the
rotation profile lead to the suggestion that the 3D shape is a
"corkscrew" filament.  These results, together with results in other
regions, suggest evolution toward higher densities as rotating
filaments shed angular momentum. Thus, magnetic fields may be an
essential feature of high-mass (M\,$\sim\,10^5$\,\msun) cloud filament
evolution toward cluster formation.
\end{abstract}


\section{Introduction}
\label{sect:Intro} 

\begin{figure*}
    \begin{minipage}{\textwidth}
    \centering
    \includegraphics[width=\textwidth]{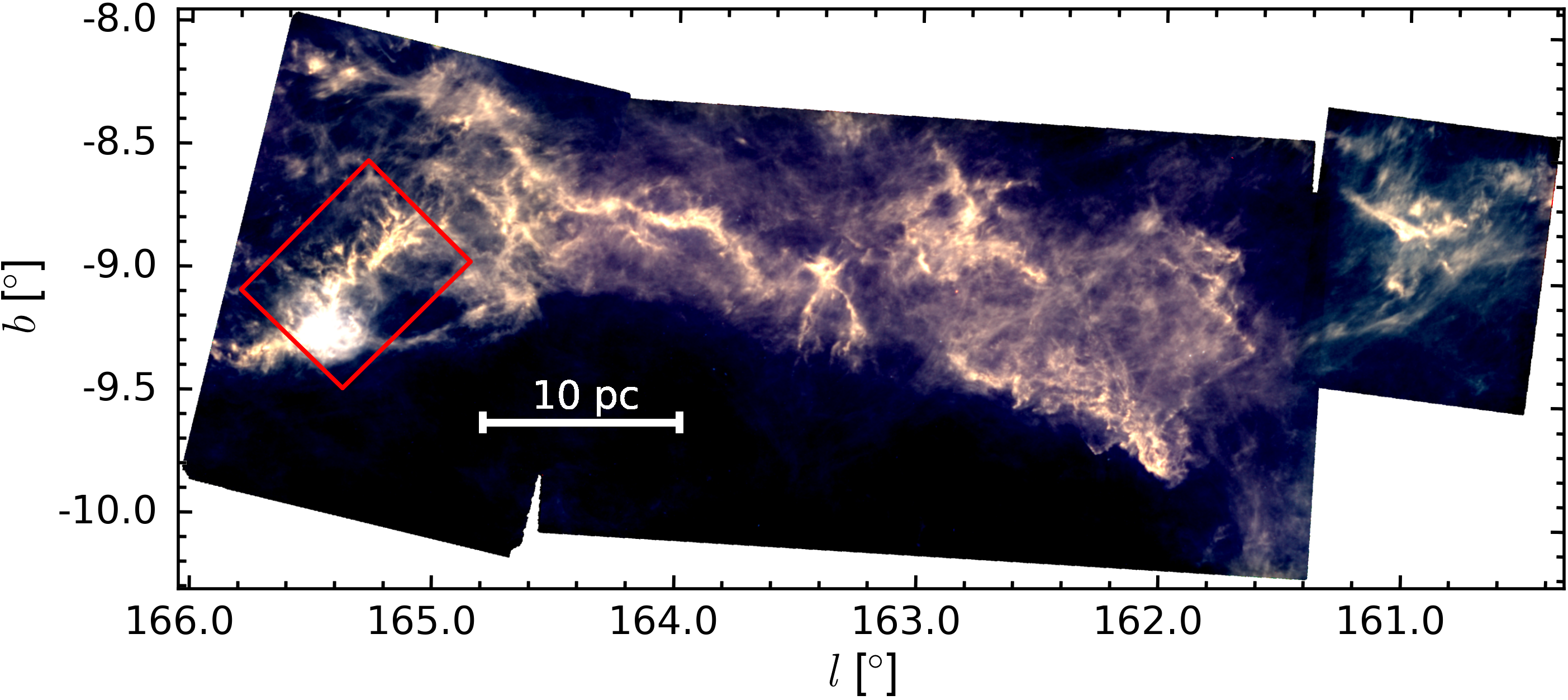} 
    \caption{False color \herschel 250~\mum (blue), 350~\mum (green)
      and 500~\mum (red) image of the California Molecular Cloud
      (CMC).  The red box indicates the approximate portion of the
      L1482 filament that we analyze in this study.  Note the
      elongated overall structure and the twisting filament spines.
      In particular, the L1482-South filament, located just to the
      north of the bright nebulosity produced by the B-star
      LkH$\alpha $ 101, has a ``zig-zag'' morphology in the plane of
      the sky (see text).}
    \label{California}
    \end{minipage}%
\end{figure*}

Using the $^{12}$CO (1-0) \citet{dame01} data,
\citet{miville-deschenes16} show that the mass function of Milky Way
clouds peaks near $\rm{M} = 10^5$\,\msun.  Moreover, most of the
molecular mass is contained in high-mass clouds, with 50\% of the mass
in clouds above $\rm{M} = 8.4 \times 10^5$\,\msun.  Therefore,
characterizing the physical properties of spiny star forming filaments
in clouds near and above the $\rm{M} = 10^5$\,\msun\ regime is
essential for studying ``typical'' star formation conditions in our
Galaxy.

Meanwhile, the clouds that we can observationally access and
scrutinize in extreme detail because of their proximity to us (with
distances $\lesssim500$~pc) typically have significantly lower masses
\citep[e.g.,][]{lada10}.  The exceptions to this are the California
Molecular Cloud (CMC), Orion~A, and Orion~B, with masses
$\sim\,1 0^5$~\msun\
\citep[e.g.,][]{megeath12,fischer17,stutz15,lada09,lada10,kong15}.
Only recently identified as a separate massive cloud \citep[][and
see below]{lada09}, the properties of the CMC dense gas and young stars
are significantly less well-characterized than those of the Orion
complex.  In this paper our primary focus is on the CMC L1482 filament
(red box in Figure~\ref{California}) gas kinematics. We place these
gas measurements in the framework required for robust comparison to
recent work in Orion~A \citep[]{stutz15,stutz16,stutz18,gonzalez19} in
order to identify possible physical differences that may explain the
variations in the observed properties of filaments and their young
stars embedded in the gravitational potentials of
M$\,\sim\,10^5$\,\msun\ clouds.

The CMC, named after its proximity to the California Nebula, was first
thought to be part of the Taurus-Auriga complex
\citep[e.g.,][]{ungerechts87,herbig04,andrews08}. It was identified as
a separate region just over a decade ago by \citet[][]{lada09}. They
found this structure to be at a much larger distance (450~pc) compared
to the nearby Taurus-Auriga (150~pc) and Perseus (240~pc) clouds.
\citet[][]{lada09} noted that the CMC had a similar mass
($\sim$10$^5$~\msun) and filamentary morphology as Orion~A, both being
nearby giant molecular clouds (GMCs).

  Another important similarity between the two was presented in
  \citet[][]{tahani2018}.  Using Faraday rotation measurements they
  found that in both the CMC and Orion~A the magnetic field flips its
  line-of-sight direction from one side of the filament to the other.
  They interpreted these results as a possible indicator of a helical
  field morphology on larger (cloud) scales \citep[but see below and
  ][for results in Orion~A]{tahani2019}.  However, the exact 3D
  magnetic field morphology in filaments (including Orion~A and the
  CMC) remains the subject of ongoing debate
  \citep[e.g.][]{heiles1987, heiles1997, matthews2000,
    fiege20a,falgarone01, hennebelle2003,stutz16,
    schleicher2018,reissl18a,tahani2019,law19,law20,reissl20}. For
  example, one suggested field morphology that could account for the
  directional flip in the field is a ``bow''-type morphology
  \citep[see e.g.,][]{inoue2018,reissl18a,
    li_klein2019,tahani2019,gomez18}. However, the simultaneous
  detection of filament rotation may be incompatible with this
  particular field geometry.  Thus, the detection of rotation in
  filaments may present one avenue to distinguish between plausible
  magnetic field geometries.  Moreover, geometries such as helical or
  toroidal may play a crucial role in filament formation. They prevent
  expansion, maintaining the filamentary structure, and may also
  influence the formation of clumps, thereby impacting the star
  formation within 
  \citep[e.g.,][]{uchida91,fiege20a,fiege20b,contreras13,stutz16}.

  More broadly, the detection of rotation in filaments (see e.g.
  \citealt{gonzalez19} for a possible rotational signature in the
  Orion~A Integral Shaped Filament) presents the opportunity to
  address the angular momentum evolution of massive star and cluster
  forming systems \citep[e.g.][]{motte2017, kong19} in the presence of
  potentially coherent and smoothly varying magnetic fields. It is also 
  essential to connect large and small scale field properties in
  star-forming systems, as will be facilitated in the near future with
  the Prime-cam instrument \citep[][]{primecam} on the Fred Young
  Submillimeter Telescope \citep[FYST, formerly CCAT-prime; ][]{ccat}.

In the southeast of the CMC we find L1482, which is one of the most 
massive filaments in the CMC and contains about 100 YSOs
\citep[][]{lada09,kong15,lada17}. L1482 hosts the reflection nebula
NGC 1579, composed of a young stellar cluster and LkH$\alpha $ 101,
the only massive (B type) star in the region \citep[e.g.,][]
{herbig04,andrews08}. Thus the CMC is mostly unaffected by massive-star 
feedback.  By scrutinizing millimeter-wave gas lines, \citet[][]{li2014} 
found that the filament structure is most likely coherent, presenting gas 
velocity gradients that they interpreted as possible inflows feeding the 
stellar cluster in the northern portion of L1482.  

As described above, the CMC and Orion~A share several important
physical attributes.  They both have a similar mass, a similar
elongated shape, both contain twisting and winding filaments, and both
have comparable magnetic field geometry (along the line-of-sight component)
and strengths as probed by Faraday rotation \citep[][]{tahani2018}. 
However, despite these
similarities, their respective dense gas fraction \citep[][]{lada09}
and YSO content are strikingly different.  In Orion~A there are more
than 3000 YSOs \citep[][] {megeath12}, whereas the CMC only has about
177 YSOs \citep[][]{lada17}.  Thus, the CMC is sometimes referred to
as a ``sleeping giant'' \citep[][]{lada17}.  As this name clearly
implies, there is a plausible evolutionary progression between the two
clouds.  Hence, assuming that the differences can be explained by the
CMC being in an earlier evolutionary state at the same mass, the CMC
may provide a window into the initial stages of filament evolution and
star-cluster formation in GMCs, before reaching the typical yet
extreme embedded cluster conditions such as those present in the Orion
Nebula Cluster and other {\it bona fide} (and hence more distant)
protoclusters.

\begin{figure*}
    \begin{minipage}{\columnwidth}
    \raggedright
    \includegraphics[width=\columnwidth]{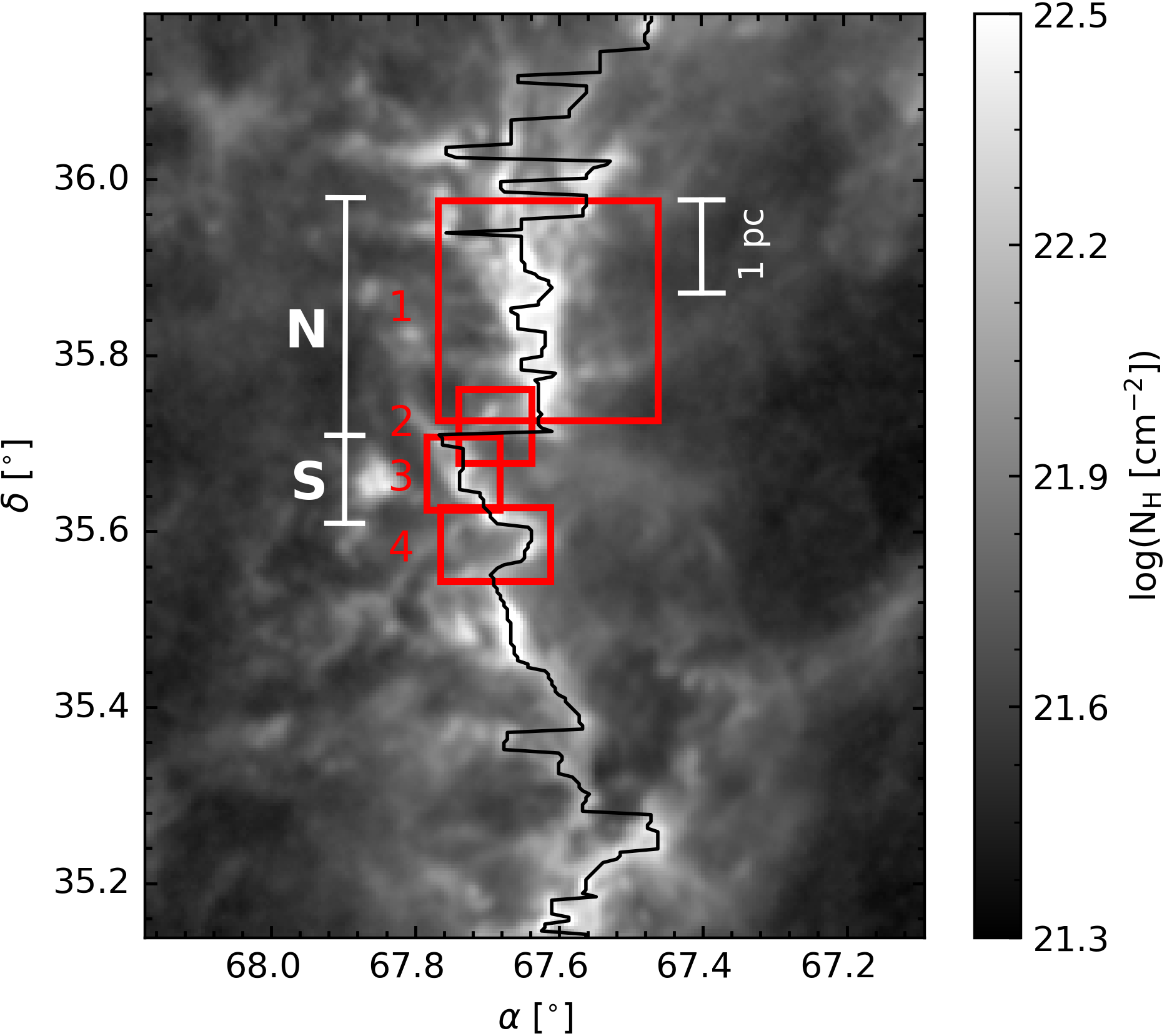} 
    \end{minipage}
    \begin{minipage}{\columnwidth}
    \hspace*{0.8cm}
    \includegraphics[width=\columnwidth]{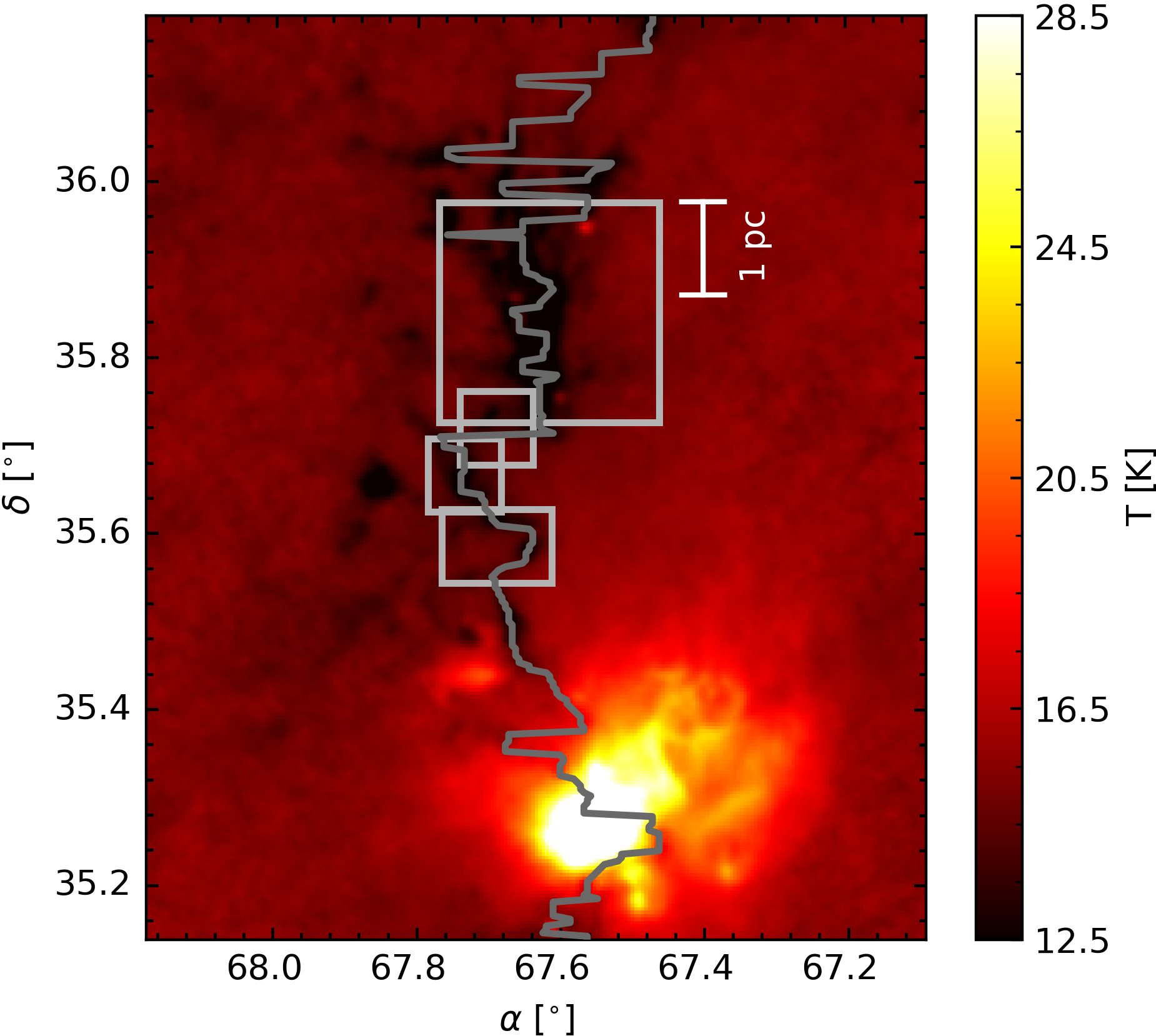} 
    \end{minipage}
    \caption{\nh (left panel) and temperature (right panel) maps of
      L1482. The black (left) and grey (right) curves represent the
      \nh ridgeline.  The red (left) and grey (right) boxes show the
      fields observed with the IRAM 30~m radio telescope.  In the left
      panel we indicate the extents of the North
      ($\delta = 35.71\degree \rightarrow 35.98\degree$) and South
      ($\delta = 35.71\degree \rightarrow 35.61\degree$) regions (see
      \S~\ref{subsect:pv_diagram}). }
    \label{fig:nh_temp}
  \end{figure*}
  
  The gas gravitational potential is a crucial parameter in a gas mass
  dominated system, such as L1482, because it sets the overall
  dynamics of the system \citep[e.g.,][]{stutz16}. Hence, it forms the
  basis for any study addressing gas (and stellar) kinematics.  In
  this framework, after presenting the data
  (\S~\ref{sect:observations}), the first parameter that we must
  determine is the distance to the system using \gaia (\S
  \ref{sect:dist}). We then estimate the gas gravitational potential
  and field based on the \herschel and \planck mass map
  (\S~\ref{sect:mass_analysis}). This sets the stage to analyze the
  gas motions (\S~\ref{sect:gas-iram}) with the goal of characterizing
  the physical state, specifically highlighting the detection of
  filament rotation of L1482-South (\S~\ref{sub:vel_grad}). We discuss
  these results in \S~\ref{sect:discussion} and conclude in
  \S~\ref{sect:conc}. 

\begin{center}
  \begin{deluxetable}{l l l l l}
    \tablecaption{Fields observed with IRAM 30m radio telescope.}
    \medskip
    \tablehead{
      \colhead{Field} & 
      \colhead{$\alpha$ [ \degree ]} & 
      \colhead{$\delta$ [ \degree ]} & 
      \colhead{size [$^{\prime\prime}\times^{\prime\prime}$]} & 
      \colhead{Observation Date}}
    \startdata
    1 &    67.62                 & 35.85                         & 900x900           &  2018 Jun 21-23 \\
     &                          &                               &                   &  2018 Jul 16    \\ 
     2 &    67.69                 & 35.72                         & 300x300           &  2019 Jan 07    \\
     3 &    67.73                 & 35.67                         & 300x300           &  2019 Jan 07    \\
     4 &    67.66                 & 35.59                         & 450x300           &  2019 Jan 07    \\
     \enddata
     \tablecomments{Central field coordinates, sizes, and
       observation dates. See Figure~\ref{fig:nh_temp}.}
     \label{table:fields}
   \end{deluxetable}
 \end{center}
 
\section{Observations}
\label{sect:observations} 

\subsection{\herschel N$_{ H}$ and T maps}\label{sect:herschel}

We use the publicly available \herschel pipeline data from the
\citet[][]{calidr} program.  These observations were made using the
PACS \citep[][]{pog10} and SPIRE \citep[][]{Griffin-spire} cameras in
parallel mode at 160~\mum, 250~\mum, 350~\mum, and 500~\mum with beam
sizes of 11.8$\arcsec$, 18.2$\arcsec$, 24.9 $\arcsec$, and
36.3$\arcsec$, respectively. Well calibrated flux data are crucial
when estimating the column density (\nh) and temperature (T) maps of a
cloud. Here we use the \citet{abreu17} method to improve the \herschel
absolute calibration by using the Planck all-sky dust model and
combining the two datasets in Fourier space. We refer the reader to
\citet[][]{abreu17} for more details.  

After combining the \herschel and \planck emission maps, we convolve
the data to the 500~\mum resolution. The convolved images are then
re-gridded to the same pixel scale of 14$\arcsec$ ($\sim0.035$~pc at
$D$ = 511~pc, see \S~\ref{sect:dist}). We then extract a spectral
energy distribution (SED) for each pixel
\citep[e.g.,][]{abreu17,stutz15,laun13,stutz13,stutz10}.  This SED is
fit with the modified black-body (MBB) function of the form:
\begin{eqnarray}
    S_{\nu} & = & \Omega B_{\nu}(\nu,T_d)(1-\exp^{-{\tau(\nu)}}),
    \label{eq:mbb}
\end{eqnarray}
where $\Omega$ is the beam solid angle, $B_{\nu}(\td)$ is the Planck
function at a dust temperature $\td$, $\tau(\nu)$ is the optical depth
at a frequency $\nu$. Here
$\tau(\nu)$~=~$N_{\rm{H}} m_{\rm{H}} R_{gd}^{-1}\kappa(\nu)$, where
$N_{\rm{H}}$~=~$2\times N({\rm{H_2}}) + N(\rm{H})$ is the total
hydrogen column density, $m_{\rm{H}}$ is the mass of the hydrogen
atom, $\kappa(\nu)$ is the dust opacity, and $R_{gd}$ is the
gas-to-dust mass ratio, which is assumed to be 110
\citep[][]{sodroski97}.  We use the dust opacities from
\citet[][]{ossen94} Table~1, column~5. \citet[][]{stutz13},
\citet[][]{laun13}, and \citet[][]{lombardi14} discuss the systematic
uncertainties produced by the model.  In Figure~\ref{fig:nh_temp} we
show the resulting \nh (left panel) and T$_{\rm{d}}$ (right panel)
maps.  We obtain similar maps as those presented in \citet{lada17}.
 
\begin{center}
    \begin{deluxetable*}{l c c c l c c c c}
    \tablecaption{IRAM 30~m observational parameters.}
    \tablehead{
    \colhead{Tracer} & 
    \colhead{Frequency} & 
    \colhead{HPBW} & 
    \colhead{$\Delta$ V} & 
    \colhead{Backend} & 
    \multicolumn{4}{c}{\ \ \ Mean Noise ($\sigma_{\rm{rms}}$) [K]} \\
    \colhead{} & 
    \colhead{\  [GHz]} & 
    \colhead{["]}  & 
    \colhead{[$\kms$]} &        
    \colhead{} &  
    \colhead{Field 1} & 
    \colhead{Field 2} & 
    \colhead{Field 3} & 
    \colhead{Field 4} }
    \startdata
        \co(1-0)               & 109.782               & 22.41         & 0.133                & FTS              &  0.26    & 0.20      &  0.20    & 0.21 \\
        \nhp(1-0)              & \, 93.173                & 26.40         & 0.157                & FTS              &  0.18    & 0.12      &  0.11    & 0.11 \\
                               &                       &               & 0.063                & VESPA            &  0.22    & 0.19      &  0.18    & 0.19 \\
        HNC(1-0)               & \, 90.663                & 27.13         & 0.161                & FTS              &  0.19    & 0.13      &  0.12    & 0.13 \\
        \hco(1-0)              & \, 89.188                & 27.58         & 0.164                & FTS              &  0.16    & 0.12      &  0.12    & 0.12 \\
    \enddata
    \label{tracers_table}
    \tablecomments{$\Delta$V is the spectral resolution for the
      respective observation. The last four columns show the noise
      levels of the data.}
    \end{deluxetable*}
\end{center}

\subsection{IRAM 30~m molecular line data}
\label{sect:iram} 

We mapped the \co (1-0), \nhp (1-0), \hco (1-0) and HNC (1-0)
molecular lines with the IRAM~30~m telescope with the primary goal of
measuring the gas kinematics. We choose \co(1-0) (109.782~GHz) as our
main tracer with the goal of obtaining measurements of the kinematics
of the gas. We also measure \nhp(1-0) (93.173~GHz), a high-density gas
tracer that enables us to observe the spine of the filament
\citep[e.g.][]{caselli2002,tafalla2002,lippok13} when detected. We
also include \hco(1-0) (89.188~GHz) and HNC(1-0) (90.663~GHz) in our
observations for mapping filament motions (such as rotation and
infall) that may be present in L1482.

\begin{figure}
    \begin{minipage}{\columnwidth}
    \includegraphics[width=\columnwidth]{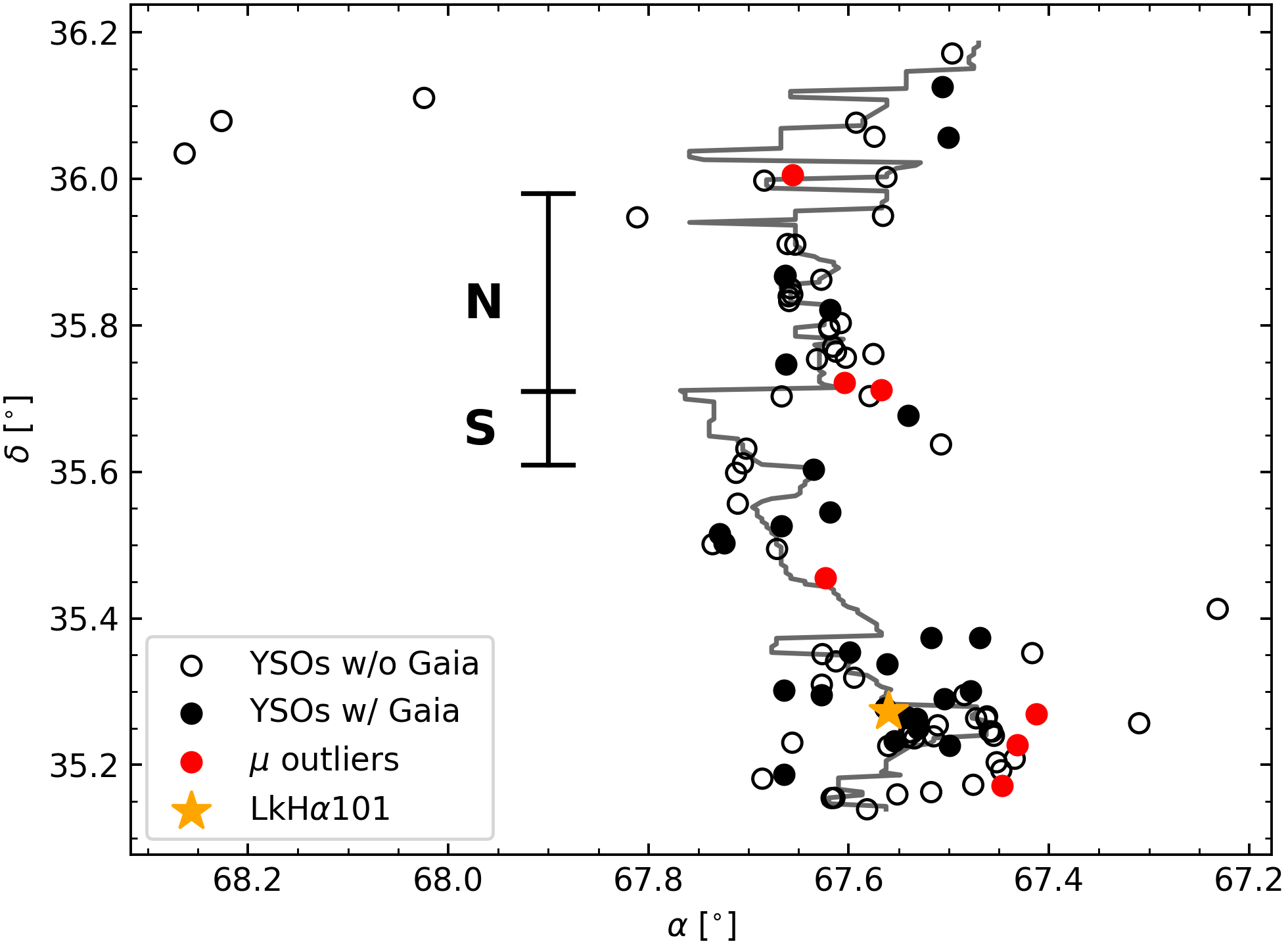}
    \caption{L1482 YSOs from \citet[][]{lada17}. We show YSOs with
      \gaia counterparts as filled circles (34 sources); of these,
      proper motion outliers are indicated in red (7 sources), with
      the remainder shown in black (see \S~\ref{sect:dist} and
      Figure~\ref{pm:sample}).  YSOs without \gaia counterparts are
      shown as black open circles (66 sources). We show the B-star
      LkH$\alpha$\,101 \citep[][]{herbig04} as an orange star. The \nh
      ridgeline is represented as a grey line (see
      \S~\ref{sect:mass_analysis}). Vertical black lines indicate the
      extent of the North (\textbf{N}) and South (\textbf{S}) regions
      (see \S~\ref{subsect:pv_diagram}).}
    \label{fig:sources}
    \end{minipage}%
\end{figure}

We use the EMIR receiver with the Fast Fourier Transform Spectrometers
(FTS) backend for all tracers. For \nhp the Versatile Spectrometric
and Polarimetric Array (VESPA) backend is also used.  All the
observations were made using On-The-Fly (OTF) mapping mode.  The
central coordinates, sizes, and observation dates are shown in
Table~\ref{table:fields}. In Table~\ref{tracers_table} we list the
spectral resolution, half power beam width (HPBW) and noise levels of
our observations.  We combine the fields and reduce the data of all
tracers using the CLASS package from
GILDAS\footnote{\href{http://www.iram.fr/IRAMFR/GILDAS}{http://www.iram.fr/IRAMFR/GILDAS}},
software developed by the Institute de Radioastronomie Millimétrique
(IRAM). Figure~\ref{fig:nh_temp} shows the location of our fields of
observation. Field~1 covers the northern portion of L1482. Field 2
covers the ``transition'' zone between the North and the South. Fields
3 and 4 cover the southern portion of L1482.

We also include the $^{13}$CO\,(2-1) data from \citep{kong15}.  These
data cover the L1482 filament as a whole and were observed with the
10~m Heinrich Hertz Sub-millimeter Telescope (SMT) on Mount Graham,
Arizona.  These data have a 35$\arcsec$ beam and a spectral resolution
of 0.15~$\kms$.  We refer the reader to \citet{kong15} for further
details.

\subsection{\gaia-detected \spitzer Young Stellar Objects}
\label{sect:gaia}

We use \gaia DR2 data \citep[][]{gaiadr2} crossmatched with the Young
Stellar Object (YSO) catalog from \citet[][]{lada17}.  In
\S~\ref{sect:dist} we obtain a robust sample for astrometric analysis
based on the parameters of \gaia-detected YSOs.

\section{\gaia distance to L1482}
\label{sect:dist}

Most of the YSOs in the \citet[][]{lada17} catalog are located in projection
on the filament (Figure \ref{fig:sources}). We therefore infer that most YSOs 
are embedded in the filament, so that their parallaxes provide a good estimate 
of the gas filament distance \citep[][]{stutz18b}.

We begin by constraining the region of our analysis to L1482, or
66.99\degree~$<$~$\alpha$~$<$~68.27\degree and
35.13\degree~$<$~$\delta$~$<$~36.18\degree. We correct the zero-point
offset in the \gaia parallaxes based on \citet[][]{zinn19}. For the
crossmatch we set the maximum separation limit to $0.5\arcsec$,
  since all Gaia-detected sources are within that limit.  We accept
  YSOs with G\,$\leq\,19$\,mag \citep[][]{lindegren2018} and positive
  parallax measurements in the \gaia catalog. Based on these criteria
  we match 34 out of 100 total YSOs in the area of interest.  See
  Figure~\ref{fig:sources} for the YSO locations relative to the \nh
  ridgeline and IRAM~30~m observations.

\begin{figure}
    \begin{minipage}{\columnwidth}
    \centering
    \includegraphics[width=\columnwidth]{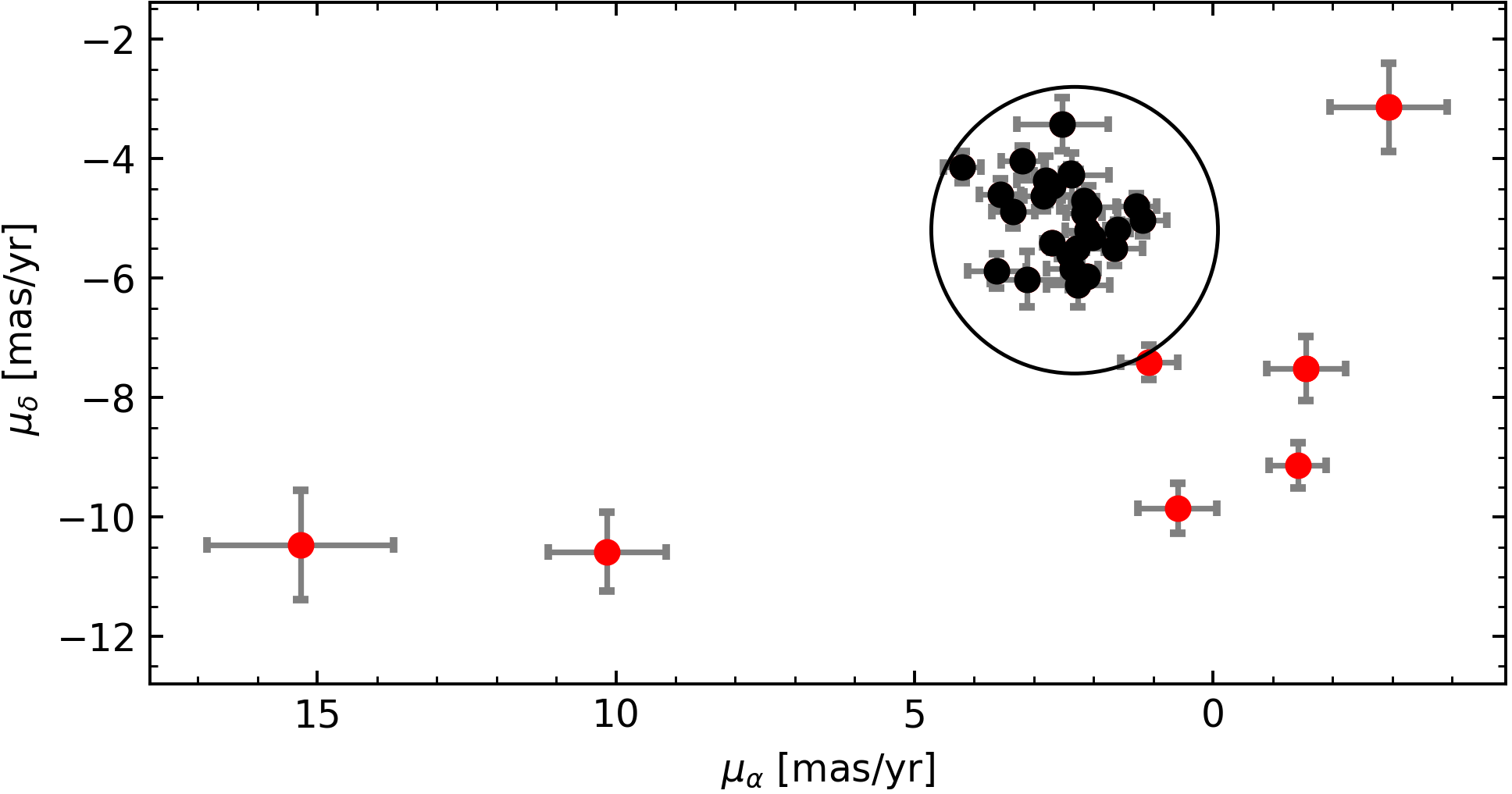}
    \caption{Proper motion ($\mu$) distribution of our
      sample of YSOs.  Proper motion outliers (shown in red) have
      $\mid\Delta\mu\mid$\,$>$\,2.4\,mas\,yr$^{-1}$ from
      the median.  See Figure~\ref{fig:sources} for the spatial
      distribution of these YSOs compared to the filament
      ridgeline.}
    \label{pm:sample}
    \end{minipage}
\end{figure}

Figure \ref{pm:sample} shows the resulting ``raw" proper motion
($\mu_\alpha$ and $\mu_\delta$) distribution of the initial 34 YSOs
that fulfill the above conditions. Here we observe seven outliers with
$\mid\Delta\mu\mid$~$>$~2.4~mas~yr$^{-1}$ from the median
of the distribution.  As noted above, these outliers are located in
projection on the filament (Figure~\ref{fig:sources}). Some of these
sources have elevated astrometric excess noise in the \gaia catalog,
while others have magnitudes near the G\,=\,19\,mag threshold, which
may worsen the reliability of their properties. We conclude that the
$\mu$ outliers show indications of corrupted astrometry,
possibly due to variability, nebulosity, and/or binarity (see below).  We
therefore exclude these from subsequent analysis.  We obtain a final
sample of 27 YSOs with reliable \gaia DR2 measurements.

\begin{figure}
  \begin{minipage}{\columnwidth}
    \centering
    \includegraphics[width=\columnwidth]{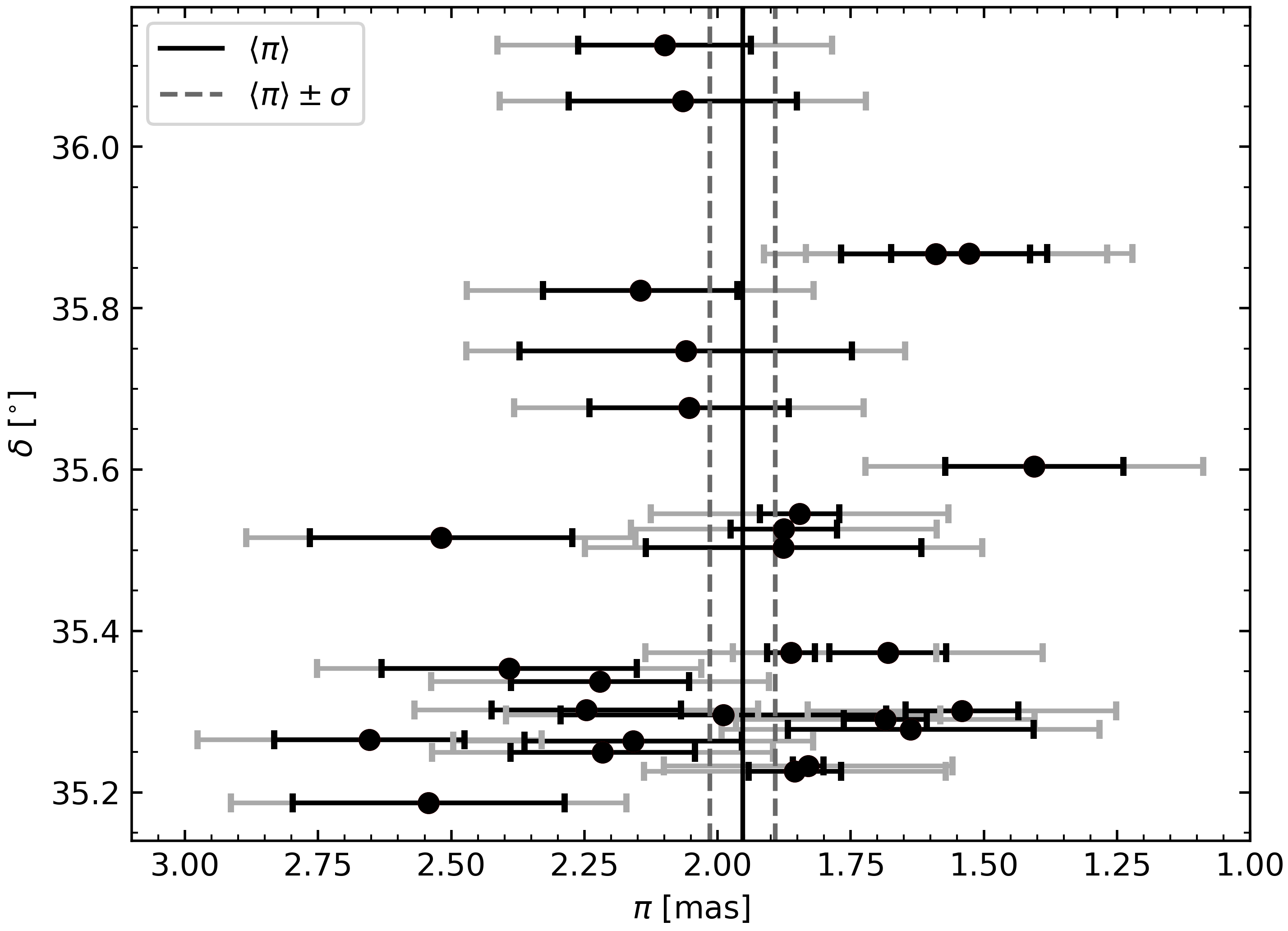}
    \caption{Parallax distribution of the 27 YSOs from \citet{lada17}
      with robust \gaia astrometry (see text).  The black points and
      black error bars show the standard \gaia catalog $\pi$ and
      $\sigma(\pi)$ values.  The grey error bars include the error
      floor $\epsilon$ calculated to enforce $\chi^2 = \rm{dof}$, see
      \S~\ref{sect:dist}.  Using the grey error bars, the
      error-weighted mean parallax and the standard error are
      $\langle\pi\rangle~=~1.956\,\pm\,0.064$~mas, shown with vertical lines in the
      plot, which corresponds to $D~=~511^{+17}_{-16}$~pc. This
      value is in good agreement with the \citet[][]{zucker2019}
      distance.}
    \label{fig:parallax}
  \end{minipage} 
\end{figure}

In Figure~\ref{fig:parallax} we show the parallax distribution of
these 27 YSOs.  We see that the intrinsic dispersion in the data
cannot be fully accounted for by the uncertainties, suggesting that
the uncertainties are under-estimates of the true uncertainties. To
correct for this we first find the mean value of the parallax,
$\mathbf \langle\pi\rangle$, the mean error, and resulting
$\chi^2 = \sum_{i=1}^{N}(\pi_i-\langle\pi\rangle)^2/\sigma_i^2$,
where $\pi_i$ and $\sigma_i$ are the Gaia-reported values. Using the 
formal errors, we obtain $\langle\pi\rangle = 1.98 \pm 0.17\,$mas 
and a corresponding $\chi^2\approx147$.  This value is large 
compared to the expected range $\chi^2 = 26\pm\sqrt{52}$ 
for a $\chi^2$ distribution with $N-1=26$ degrees of freedom 
\citep[dof; see][]{gould04}.  It is plausible that the true 
\gaia errors are larger than the reported formal uncertainties in 
the region of the filament. Measurements can be adversely affected 
by nebulosity, crowding, binaries, and differential extinction to 
a substantially greater degree than for field stars
\citep[][]{arenou18,lindegren2018,gaia18,zinn19,penoyre20,rao20}.  We
therefore define the augmented reported errors as
$\sigma_i^\prime = \sqrt{\sigma_i^2 + \epsilon^2}$, where $\epsilon$
is an error floor.  We choose $\epsilon$ to enforce $\chi^2=26$ using
the renormalized errors.  We find $\epsilon=0.29$ mas. We measure the
error weighted mean parallax and its standard error on the mean of
$\langle\pi\rangle = 1.956 \pm 0.064\,$mas for the parallax,
corresponding to $D = 511^{+17}_{-16}\,$pc for the distance.  The
measured distance is consistent with the \citet[][]{zucker2019}
distance of 524$^{+11}_{-7}$~pc.

To test for a possible signature of inclination in the filament, we
fit the $\pi$ values as a function of $\delta$ \citep[][]{stutz18b}.
We find no robust trend, in agreement with the visual impression from
Figure~\ref{fig:parallax}.  Given the small number of data points and
their corresponding errors, this is not a strong constraint, and hence
the filament could still have significant inclination.

\begin{figure}
    \begin{minipage}{\columnwidth}
    \centering
    \includegraphics[width=\columnwidth]{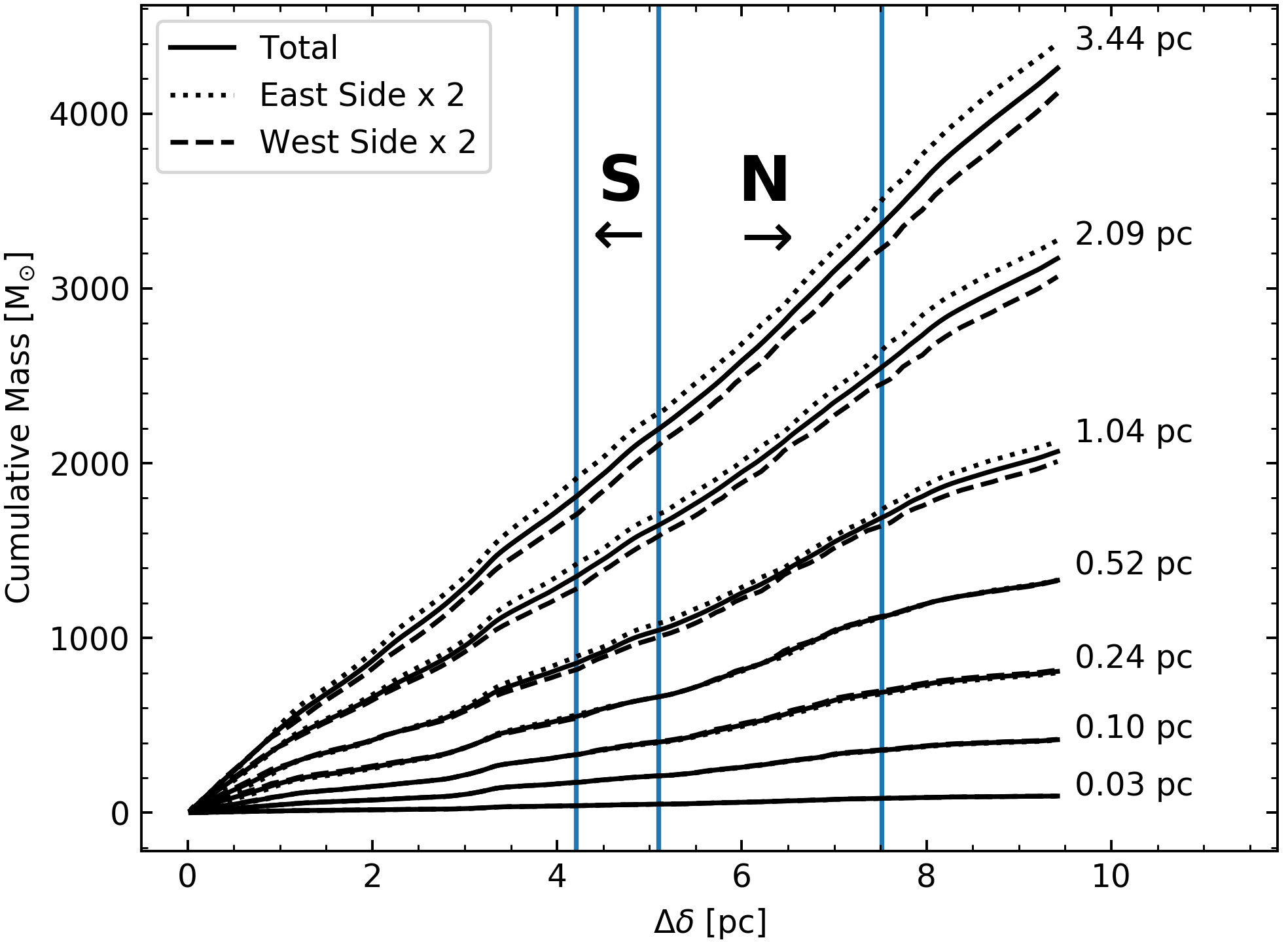}
    \caption{Cumulative mass distribution along the L1482 filament.
      Different curves indicate different projected radii or impact
      parameters (see values on the r.h.s.)  from the \nh
      ridgeline. Both the east (dotted black lines) and west (dashed
      black lines) profiles are multiplied by 2 for visual comparison.
      At all radii the mass as a function of $\delta$ grows
      approximately linearly.  The declination offset, $\Delta\delta$,
      starts at the bottom of Figure~\ref{fig:nh_temp} and runs from
      $\delta\,=\,35.14\degree \rightarrow \delta\,=\,36.18\degree$.
      The North and South regions (\S\ref{subsect:pv_diagram}) are
      indicated with vertical blue lines.}
    \label{cumulative-mass} 
    \end{minipage} 
\end{figure}

\section{Mass, line-mass, gravitational potential, and acceleration}
\label{sect:mass_analysis} 

Here we present the mass distribution analysis based on the \herschel
\nh map (see \S~\ref{sect:herschel}). In order to carry out this
analysis, we start with the identification of the \nh ridgeline.
We use the \nh map (Figure~\ref{fig:nh_temp}) to identify the ridgeline 
of peak gas surface density as a function of declination $\delta$. 
Using the \nh ridgeline, we separate the map into east and west sides.  
In Figure~\ref{cumulative-mass} we show the cumulative mass along $\delta$ 
at different projected radii from the \nh ridgeline. We show the extents 
of the North and South regions (see \S~\ref{subsect:pv_diagram}).

From this figure we appreciate two important features of the mass
distribution. First, at each projected radius, the mass increases
almost linearly without jumps along $\delta$. Hence, the cumulative
mass distribution is, to first order, only dependent on the projected
radius. Second, the distributions are mostly symmetric on each side of
the filament, simplifying the geometry of the system.  Combined, these
features allow us to extract an average cumulative line-mass (M/L)
profile for the structures as a whole \citep[][]{stutz16}.

\begin{figure*}
  \centering
  \includegraphics[width=18cm]{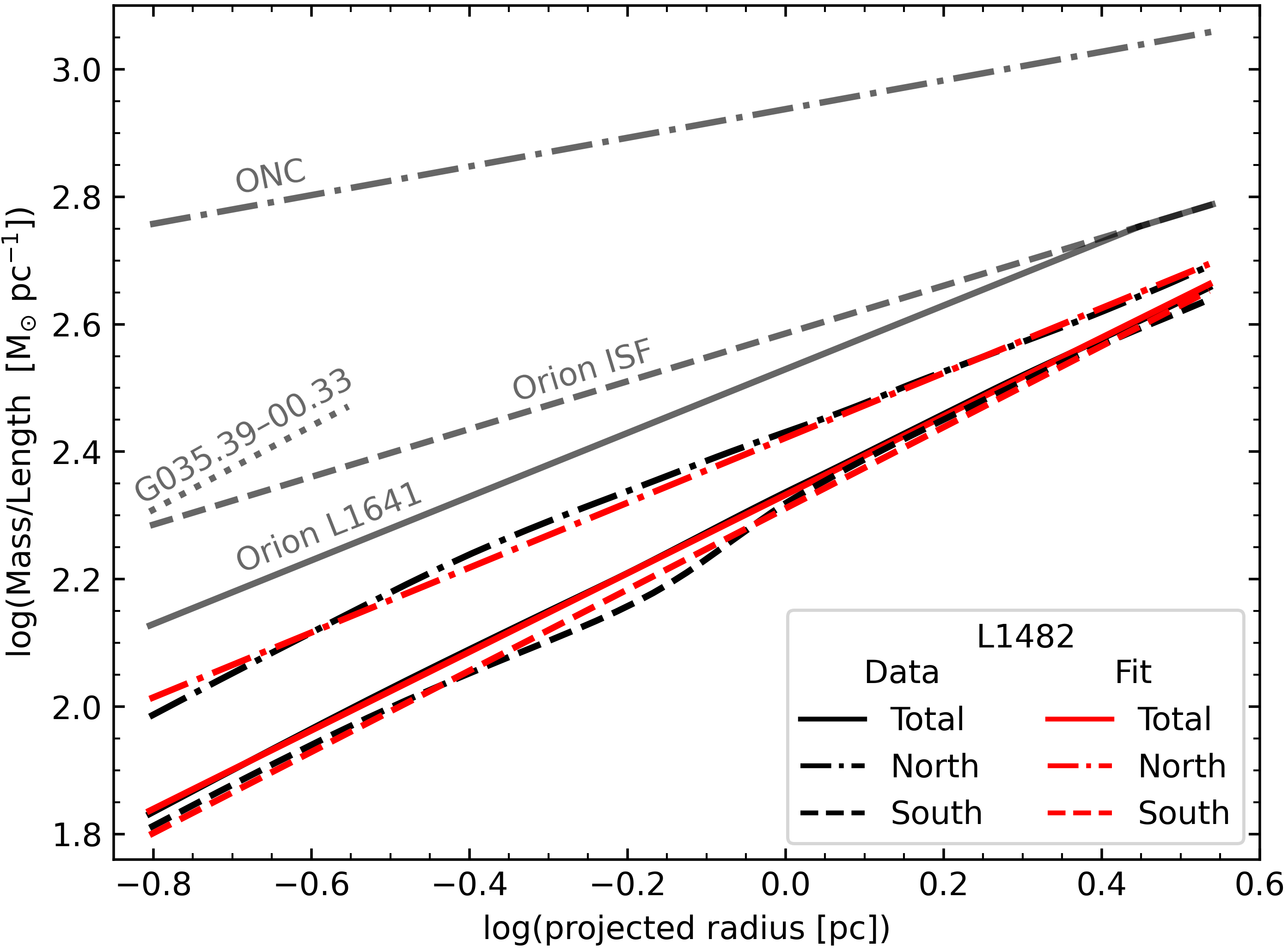}
  \caption{Enclosed line-mass (see Equation~\ref{eq:power_law} and
    Table~\ref{table:mass_param}) versus projected radius from the \nh
    ridgeline for the L1482 filament (black curves) and best-fit power
    law (red curves).
    We include the profiles for the North and South regions (see
    \S~\ref{subsect:pv_diagram} and \S~\ref{sub:mach}). For
    comparison, we show $\lambda(\rm{w})$ for Orion ISF (grey dashed
    curve; \citealt[][]{stutz16}), Orion L1482 (grey solid curve;
    \citealt[][]{stutz16}), the ONC (grey dashed dotted line;
    \citealt[][]{stutz18}), and G035 (dotted gray line; measured from
    the \citealt[][]{kainulainen13a}; see \S~\ref{sect:discussion} for
    details of this filament). This figure presents a progression from
    low to high line-mass regions, where the lowest line-mass region
    is L1482 South (dashed line) with comparatively few YSOs, and the
    highest line-mass region is the Orion Nebula Cluster \citep[ONC,
    ][]{stutz18}, with hosting the embedded star cluster by the same
    name.}
    \label{fig:enclosed-line-mass}
\end{figure*}

We show the enclosed M/L profile of L1482 in
Figure~\ref{fig:enclosed-line-mass}.  We include the profiles for the
North and South regions separately. The distributions are well approximated by a
power-law down to the resolution limit of the data. We apply a
power-law fit (i.e., red-line linear fit to the log-log representation
in Figure~\ref{fig:enclosed-line-mass}) to the filament profiles,
including the North and South regions:
\begin{eqnarray}
    \lambda_{app}(w) & = & \rm{\zeta}
                     \biggl(\frac{\textit{w}}{\pc}\biggr)^\gamma,
    \label{eq:power_law}
\end{eqnarray}
where $w$ is the plane-of-the-sky (POS) projected radius; this
expression gives the apparent line-mass distribution in the POS (see Table~\ref{table:mass_param} for the values of $\zeta$ and
  $\gamma$).  This demonstrates that the North region has a higher
line-mass profile than the South region. Meanwhile, the total
line-mass profile of L1482 is lower than those of the Orion L1641,
Orion Integral Shaped Filament (ISF), and the Orion Nebula Cluster
\citep[ONC, ][]{stutz16,stutz18}, respectively.

Because of the symmetry and radial dependence of the cumulative mass
profiles, and the fact that the line-mass profiles are
well-characterized by scale-free power laws down to the resolution
limit of the \herschel data (see Figure~\ref{cumulative-mass} and 
\ref{fig:enclosed-line-mass}), we assume a
cylindrical morphology for the filament, with axial symmetry around
the \nh ridgeline.  Below we follow the formalism presented in
\citet[][]{stutz16} and \citet[][]{stutz18} for the calculations of
the apparent plane-of-the-sky volume density, gravitational potential,
and gravitational acceleration.  Table~\ref{table:mass_param} presents
the power-law indices and normalizations for the expressions presented
below.

\begin{deluxetable*}{l c c c c c c c }
        \tablecaption{Line mass, density, gravitational potential, and
          acceleration profile parameter values.}
        \tablehead{
        \colhead{Region} &  
        \colhead{$\zeta^g$}   & 
        \colhead{$\beta^h$} & 
        \colhead{$\psi^i$}   &  
        \colhead{$\xi^j$}   &   
        \colhead{$\gamma^k$}&  
        \colhead{Projected length}&
        \colhead{Total gas mass}\\ 
        \colhead{}  &
        \colhead{{$[\msun \pc^{-1}]$}} & 
        \colhead{{$[\msun \pc^{-3}]$}} & 
        \colhead{{$[(\kms)^{2}]$}} & 
        \colhead{{$[(\kms)^{2} \pc^{-1}]$}} &
        \colhead{} &
        \colhead{{[pc]}} &
        \colhead{$[\msun]$}}
        \startdata
            L1482$^a$       &    214     & 10.2    & 1.45    &  0.89  &  0.62  & 9.4 & 4260    \\
            L1482-N$^b$     &    264     & 12.6    & 2.63    &  1.34  &  0.51  & 2.3 & 1090    \\
            L1482-S$^c$     &    205     & \, 9.6  & 1.27    &  0.81  &  0.64  & 0.9 & \, 380     \\
        \hline                                                                               
            Orion ISF$^d$   &    385     & 16.5    & 6.30    &  2.40  &  0.38  & 7.3    & 6200     \\
            Orion L1641$^d$ &    338     & 16.1    & 3.50    &  1.80  &  0.50  & 23.2 \,& $2\times10^4$     \\
            ONC$^e$         &    866     & 25.9    & 27.60 \,&  6.40  &  0.23  & 0.5    & 1300 \\
        \hline
            G035$^f$           &    682     & 31.5    & 3.96    &  2.60  &  0.66  & 2.0  & \,822\\
        \enddata
        \tablecomments{Values for the parameters derived in \S~\ref{sect:mass_analysis}.
          We include all regions shown in Figure~\ref{fig:enclosed-line-mass}.\\
        $^a$ This work, calculated over the range $35.14\degree < \delta < 36.19\degree$.\\
        $^b$ This work, calculated over the range $35.71\degree < \delta < 35.98\degree$. \\
        $^c$ This work, calculated over the range $35.61\degree < \delta < 35.71\degree$.\\
        $^d$ \citet[][]{stutz16}.\\
        $^e$ \citet[][]{stutz18}.\\      
        $^f$ Calculated using a \nh map derived from \citet[][]{kainulainen13a} over the \nhp region from \citet[][]{henshaw14}.\\
        \noindent Normalization constants:\\
        $^g$ for the M/L profile (Eq.~\ref{eq:power_law}).\\
        $^h$ for the volume density (Eq.~\ref{eq:density}).\\
        $^i$ for the gas gravitational potential (Eq.~\ref{phigrav}).\\
        $^j$ for the gravitational acceleration (Eq.~\ref{grav_acc}).\\
        $^k$ is the power-law index in the M/L and gas gravitational
        potential profiles; the power-law index for the volume density
        is $\gamma-2$ and for the gravitational acceleration is
        $\gamma-1$.}
      \label{table:mass_param}
\end{deluxetable*}

The apparent volume density is estimated as:  
\begin{eqnarray}
    \rho_{app}(r) & = & \frac{\gamma(-\gamma/2)!}{2(-\gamma/2 - 1/2)!(-1/2)!}
    \frac{\zeta}{\pc^2} \left(\frac{r}{\pc}\right)^{\gamma-2} \\
     & = & \beta\left(\frac{r}{\pc}\right)^{\gamma-2}.
    \label{eq:density}
\end{eqnarray}
The gas gravitational potential follows as:
\begin{eqnarray}
    \Phi_{app}(r) = \psi\bigg(\frac{r}{\pc}\bigg)^{\gamma}.
    \label{phigrav}
\end{eqnarray}
Given that the gravitational acceleration is defined as
$g(r)~=~-\nabla \Phi(r)$, then:
\begin{eqnarray}
    g(r)_{app} =  -\xi \bigg(\frac{\it{r}}{\pc}\bigg)^{\gamma-1}. 
    \label{grav_acc}
\end{eqnarray}
The expressions in Equations~(\ref{eq:power_law}), (\ref{eq:density}),
(\ref{phigrav}), and~(\ref{grav_acc}) represent the plane-of-the-sky
inferred quantities.  They must be multiplied by the unknown
projection factor $\cos(\theta)$, where $\theta$ is the inclination
angle of the filament relative to the plane of the sky (see \S~\ref{sect:dist}
for inclination discussion).  In Table~\ref{table:mass_param} we give
$\zeta$, $\beta$, $\psi$, $\xi$, and $\gamma$ values for all regions
in Figure~\ref{fig:enclosed-line-mass}, under the assumption that
$\cos(\theta) = 1$.

We also analyze the inner mass distribution of IRDC G035.39-00.33
(henceforth G035), located in the W48 complex at a distance of 
2.9~kpc. This cloud has a total mass of
$\sim 2\times~10^4~\rm{M}_{\odot}$, presents a filamentary morphology,
gas velocity gradients, and low star formation activity in the
northern portion \citep[][]{simon06,kainulainen13a,nguyen11,henshaw14}.  
We start with the 8\mum extinction map from \citet[][]{kainulainen13a} 
to derive a \nh map of the region. We repeat the mass distribution 
analysis described above to the subregion analyzed in \citet[][]{henshaw14}. 
See Table~\ref{table:mass_param} for the parameters.  After the ONC,
G035 has the highest line-mass of the regions shown in
Figure~\ref{fig:enclosed-line-mass} (see \S~\ref{sect:discussion}).

\section{Filament gas velocities}
\label{sect:gas-iram}

Gas radial-velocity maps provide velocity gradients, which may be
signatures of rotation, outflows, and/or infall in L1482. Measurements
of the line widths allow us to scrutinize the non-thermal velocity
dispersion.  When compared to the gas gravitational potential, we can
then evaluate the physical state of the filament
\citep[e.g.,][]{gonzalez19}.  In our analysis below, we use IRAM 30~m
data (\S~\ref{sect:iram}) and the \citet{kong15} $^{13}$CO(2-1) data
(\S~\ref{sect:iram}).
 
\begin{figure}
    \begin{minipage}{\columnwidth}
    \includegraphics[width=\columnwidth]{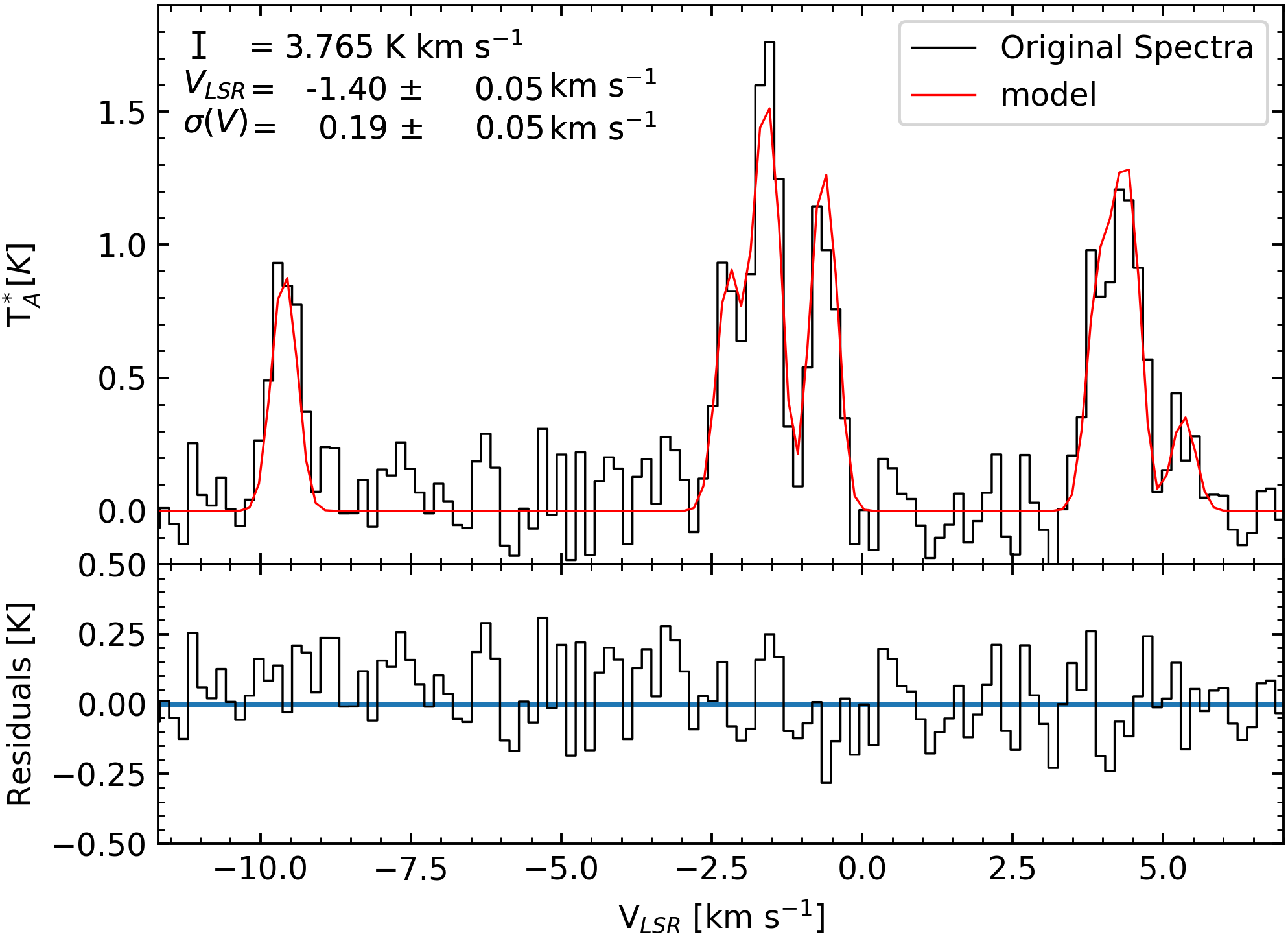}
    \caption{\nhp line fitting example for a single pixel located at
      $\alpha$~=~67.6\degree $\delta$~=~35.97\degree.  {\it Top:} The
      original spectrum (black curve) and the fitted model (red
      curve). In the top left corner we show the integrated intensity,
      the velocity centroid, and the velocity dispersion. {\it
        Bottom:} The residuals of the fit are shown in black and are
      centered at y\,=\,0, indicated with a blue line.}
    \label{n2h+_fit}
    \end{minipage}
  \end{figure}
  
\subsection{Line fitting}
\label{sub:fitting}

We employ the line-modeling python package PySpecKit
\citep[][]{ginsburg2011} to fit line profiles and remove noise from
our data. For the line fitting of \co, \hco, HNC, and \sco we use a
single Gaussian fit, included in PySpecKit. The derived parameters for
these tracers are the peak of the spectrum, velocity centroid, and
velocity dispersion. To model the hyperfine line structure of \nhp, we
use the built-in fitter ``n2hp$\_$vtau'' \citep[][]{ginsburg2011} that
adjusts multiple Gaussians to the raw spectrum. The derived parameters
are the excitation temperature, optical depth, integrated intensity,
velocity centroid, and velocity dispersion.

To set a signal-to-noise ratio (SNR) threshold for the fitter we need
to determine the SNR for each spectrum.  We create an error map by
taking the RMS ($\sigma_{\rm{rms}}$) values of the spectra in a
velocity range without gas emission. We set this velocity range from
$-40$~$\kms$ to $-10$~$\kms$ for the IRAM 30~m tracers. The mean
$ \sigma_{\rm{rms}}$ for each field is shown in
Table~\ref{tracers_table}. For \sco we set the velocity range between
$-25$~$\kms$ and $-10$~$\kms$. For \co, \hco, and HNC we find a SNR
threshold of 3.5. For both \sco and \nhp, SNR~=~3 produces good fitted
models. These values remove most of the noise without affecting the
emission from the filament.

\begin{figure}
  \begin{minipage}{\columnwidth}
    \centering
    \includegraphics[width=\columnwidth]{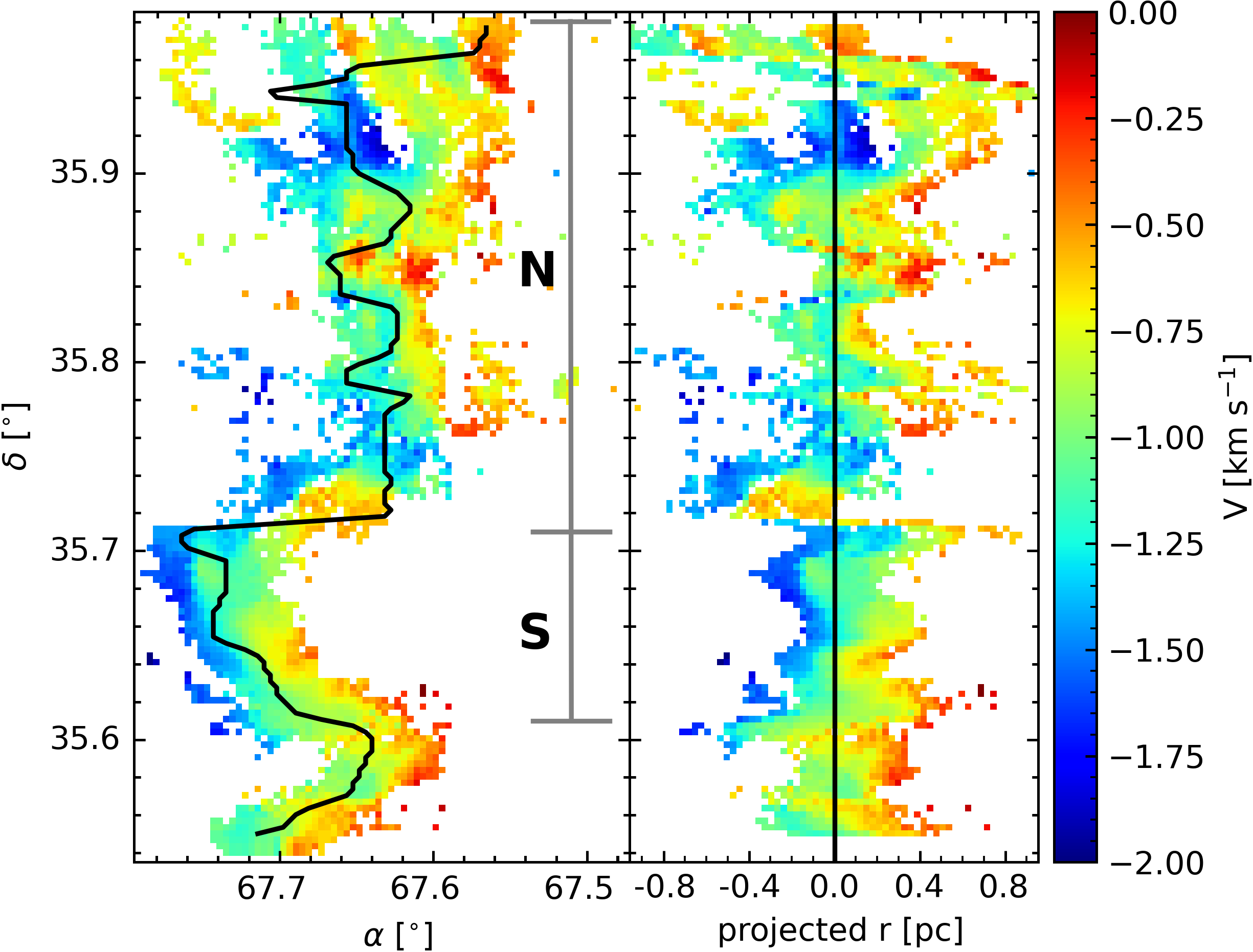}
    \caption{{\it Left:} \co first-moment (mean velocity) map. The
      black curve shows the \nh ridgeline. {\it Right:} \co
      first-moment map aligned to the \nh ridgeline (straight black
      line). Negative r values indicate the east side of the filament.
      We observe a velocity gradient (VG) perpendicular to the
      long axis of the filament, going from blue- to red-shifted
      velocities from east to west.  This rotation signature is
      particularly evident in the southern portion of the filament
      (extent indicated with the vertical grey bar in the left 
      panel), see text.}
    \label{fig:aligned_tracers}                 
  \end{minipage}
\end{figure}

The fitting process requires a user-defined starting point inside the
cube and initial guesses to fit the spectrum in this pixel. For all
tracers we use starting values based on the peak of the spectra,
first-moment map, and second-moment map values. For \nhp we also 
adopt an excitation temperature and optical depth of 6~K and 0.5
respectively. These last two parameters were selected by testing
different values until obtaining fitter result convergence. In
Figure~\ref{n2h+_fit} we show one pixel example of the \nhp modeled
cube. In the top panel we show the original spectrum (black line) and
the fit (red line). In the lower panel we show the residuals.

\begin{figure*}
    \includegraphics[width=\textwidth]{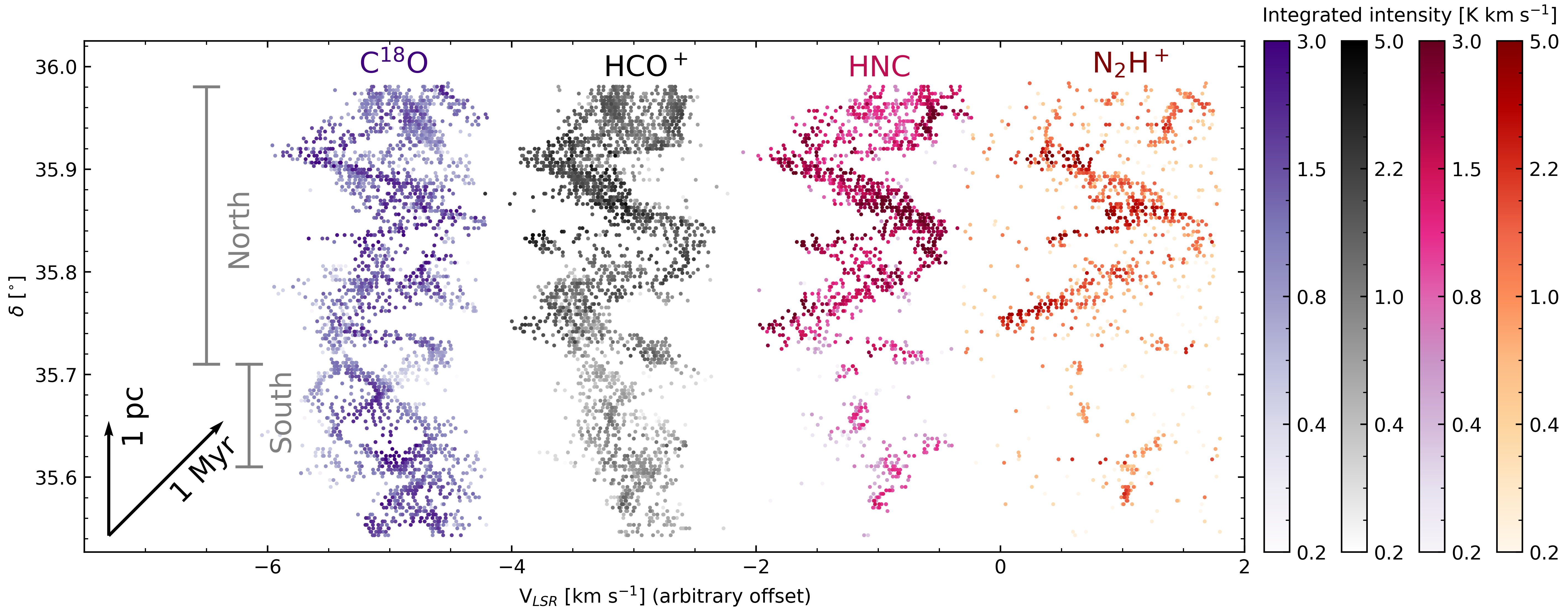}
    \caption{Intensity-weighted position-velocity (PV) diagram: \co,
      \hco and \nhp are shown with arbitrary velocity offsets of
      $-4~\kms$, $-2~\kms$, and $+2~\kms$, respectively, from HNC. The
      color bars are logarithmic in scale to enhance the visibility of
      structures with lower intensities.  The gray vertical lines
      indicate the extent in $\delta$ of the North and South regions
      defined in \S~\ref{subsect:pv_diagram}. The black vertical arrow 
      shows the spatial scale of 1~pc, and the diagonal arrow indicates 
      a timescale of 1\,\pc/$\kms$\,$\approx$\,1\,Myr.}
\label{pv}
\end{figure*}

For the single-component IRAM 30~m models we remove artifacts produced
by bad fits, located in the outer parts of the filament at low SNR.
We define SNR~=~I$_{\rm{peak}}$/$\sigma_{\rm{rms}}$, where
I$_{\rm{peak}}$ is the peak intensity of the modeled spectra and
$\sigma_{\rm{rms}}$ is derived from the original data in the same
velocity range as in the fitter. We measure the SNR in all cubes. For
consistency with the fitter procedure we remove spectra with
SNR~$<$~3.5. We obtain good results with this approach, giving us
clean models to work with.

In Figure~\ref{fig:aligned_tracers} we present the cleaned version of
the modeled \co mean line velocity (moment 1) map. In the left panel
we show the standard moment 1 map, while in the right panel we align
the map at each $\delta$ to the \nh ridgeline. From this figure it is
immediately obvious, especially in the South region, that we detect a
clear and confined velocity gradient (VG) going from
negative~$\rightarrow$~positive velocities from
east~$\rightarrow$~west, ``hugging'' the \nh ridgeline. 
Gas velocity gradients have been detected in different filamentary 
systems with different proposed origins such as filamentary rotation,
shear, gas inflow, and cloud-cloud collisions \citep[e.g.,][]{
uchida91,jimenez14,lee14,fernandez14,henshaw14}.
We study this velocity gradient in detail in \S~\ref{sub:vel_grad}.

\begin{figure}
    \begin{minipage}{\columnwidth}
    \includegraphics[width=\columnwidth]{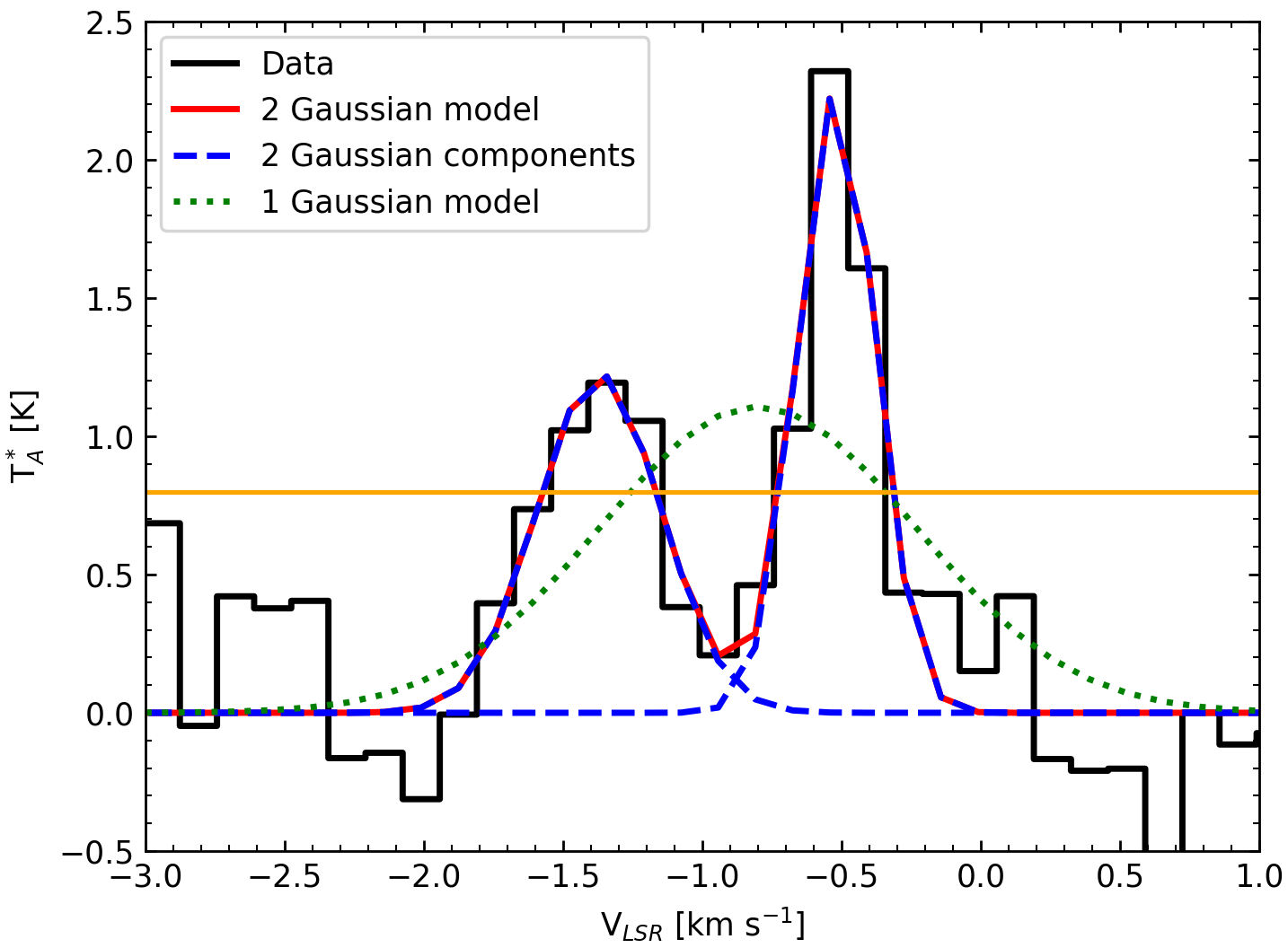}
    \caption{Double Gaussian spectrum example. The raw data is
        shown in black. The double Gaussian fit is shown in red and
        its components with dash blue lines. The single Gaussian fit
        is presented with a green dotted line. We indicate the
        intensity threshold with a solid orange line.}
\label{fig:2g_example}
\end{minipage}
\end{figure}

For \sco, \co, \hco, and HNC the line widths present large variations
along $\delta$.  Here we test whether the large velocity dispersions
are the result of fitting a double-component or more complex line
profile with a single Gaussian by using a double-component model.
From the double velocity-component fitting results, we find that most
spectra can be fitted with only one Gaussian, while the secondary
component generally has low SNR compared to the defined threshold. In
order to identify spectra with reliable double components we define a
temperature threshold. If the peaks of both components are above the
threshold we consider them as well-detected. For \co, we identify two
well fitted velocity components when we set this threshold to 0.8~K,
almost three times the noise value of Field 1 (see Table
\ref{tracers_table}).  From the above procedure, we conclude that
$\sim$13\% of the \co data contains double-components.  These are
mainly located at or near the \nh ridgeline.  When fit individually,
the components of these spectra have slightly lower line widths
compared to the single components.  The North region presents more
double-component spectra compared to the South, as we expect (see
discussion below).  In order to avoid line-width and velocity
contamination or confusion in the subsequent analysis, we remove the
pixels in which double-component spectra are detected
(see Figure~\ref{fig:2g_example}).

In short, the single component spectra dominate the spectral cubes, and
the measured velocities and the line widths of the fitting procedure agree
well with previous observations of L1482
\citep[e.g.,][]{li2014,kong15}.  The removal of the double component
spectra does not affect the results of our analysis below.  Moreover, 
we find no \co or \hco spectra exhibiting a clear blue asymmetry or an 
inverse P-Cygni like profile. Such profiles would indicate infall along the
line of sight \citep[e.g.,][]{myers2000,evans15, smith12}.  The
absence of an infall signature may be caused by the resolution of our
data.  We conclude that our procedure adopting single-component fits
is robust.  In the sections that follow we use these to analyze the
velocity structure of L1482.

\subsection{Intensity-weighted position-velocity diagrams}
\label{subsect:pv_diagram}

Using the fitter results described above, we generate
intensity-weighted position-velocity (PV) diagrams for the IRAM~30~m
data, presented at different velocity offsets in Figure~\ref{pv}. Here
we show the best-fit gas velocities as a function of $\delta$ along
the filament, weighing by the integrated line intensity.  This
technique is described in detail in \citet[][]{gonzalez19}; it removes
noise present in the traditional PV diagram method while highlighting
structure that would otherwise be muddled or invisible (see their
Figure~3 and~4).

In Figure~\ref{pv}, we see similar structures in \co and \hco across
most of the extent of the filament.  High density gas, as traced by
\nhp, is mostly not detected at $\delta \lesssim 35.71\degree$.
Moreover, at $\delta \sim 35.71\degree$ there is a discontinuity in
the filament that is coincident with a jump in the \nh map (see
Figure~\ref{fig:nh_temp} and \ref{fig:aligned_tracers}).  This jump
marks the transition between two physically distinct filament
environments.  Above this location we have a higher M/L region, which
contains more YSOs (see Figure~\ref{fig:sources}), while below the jump
we find a more confined filament with lower M/L values (see
Figure~\ref{fig:enclosed-line-mass} and
Table~\ref{table:mass_param}). We therefore divide the filament into
two subregions. The northern region encompasses
$\delta = 35.71\degree \rightarrow 35.98\degree$.  The southern region
encompasses $\delta = 35.71\degree \rightarrow 35.54\degree$.

In Figure~\ref{pv} we observe in all tracers the presence of elongated
structures with gradients.  These have slopes, given the axis ratio of
the plot, approximately consistent with $1$~Myr timescales (but see 
discussion below).  Along the filament we also identify structures that 
have an appearance consistent with wrapping or winding, perhaps most 
obvious in \co in the southern portion of the filament.  In this region 
the filament has a clear ``zig-zag'' morphology in \nh (see 
Figure~\ref{fig:nh_temp}), potentially indicating a cork-screw or 
helical-like morphology in 3D. Overall, the velocity wiggles are 
reminiscent of the structures in the Integral Shaped Filament 
(ISF) in Orion~A \citep[see Figure~4 of ][]{gonzalez19}.  

At $\delta \sim 35.83$\degree we observe a large spike in velocity
where the filament appears to have two well-separated velocity
components, most obvious in \hco, HNC, and to a lesser extent in \nhp.
The region at the east of the ridgeline in the \co first moment map
(see Figure~\ref{fig:aligned_tracers}) exhibits compact blue- and
red-shifted velocities near this location.  These alone might
plausibly indicate some YSO-associated outflow activity.  However, the
appearance of this feature, albeit at fainter levels, in \nhp
indicates that this velocity pattern persists in the denser gas, where
outflow signatures may be less likely to arise.

The feature bears some resemblance to the \nhp velocity spike observed
by \citet[][]{gonzalez19} in Orion~A at $\delta \sim 5.4$\degree, that
is, in the center of the ONC gas filament. In the present case, the
maximum velocity shift between the two \nhp loci is $\sim 1.5$~$\kms$,
while in the ONC it is $\sim 4$~$\kms$.  The \co velocity pattern
continues to the South to $\delta \sim 35.74\degree$, progressing
through a series of back-and-forth wiggles until
$\delta = 35.71\degree$, where the filament breaks and jumps over both
in projected position (in the \nh map, and see above) and in velocity,
which may bear some relation to the drastic break in the ISF near
$\delta \sim -5.5\degree$.  The general appearance of the velocity
patterns in L1482 is somewhat similar to the ISF
\citet[][]{gonzalez19}, but with smaller amplitudes.

\begin{figure}
    \begin{minipage}{\columnwidth}
    \centering
    \includegraphics[width=\columnwidth]{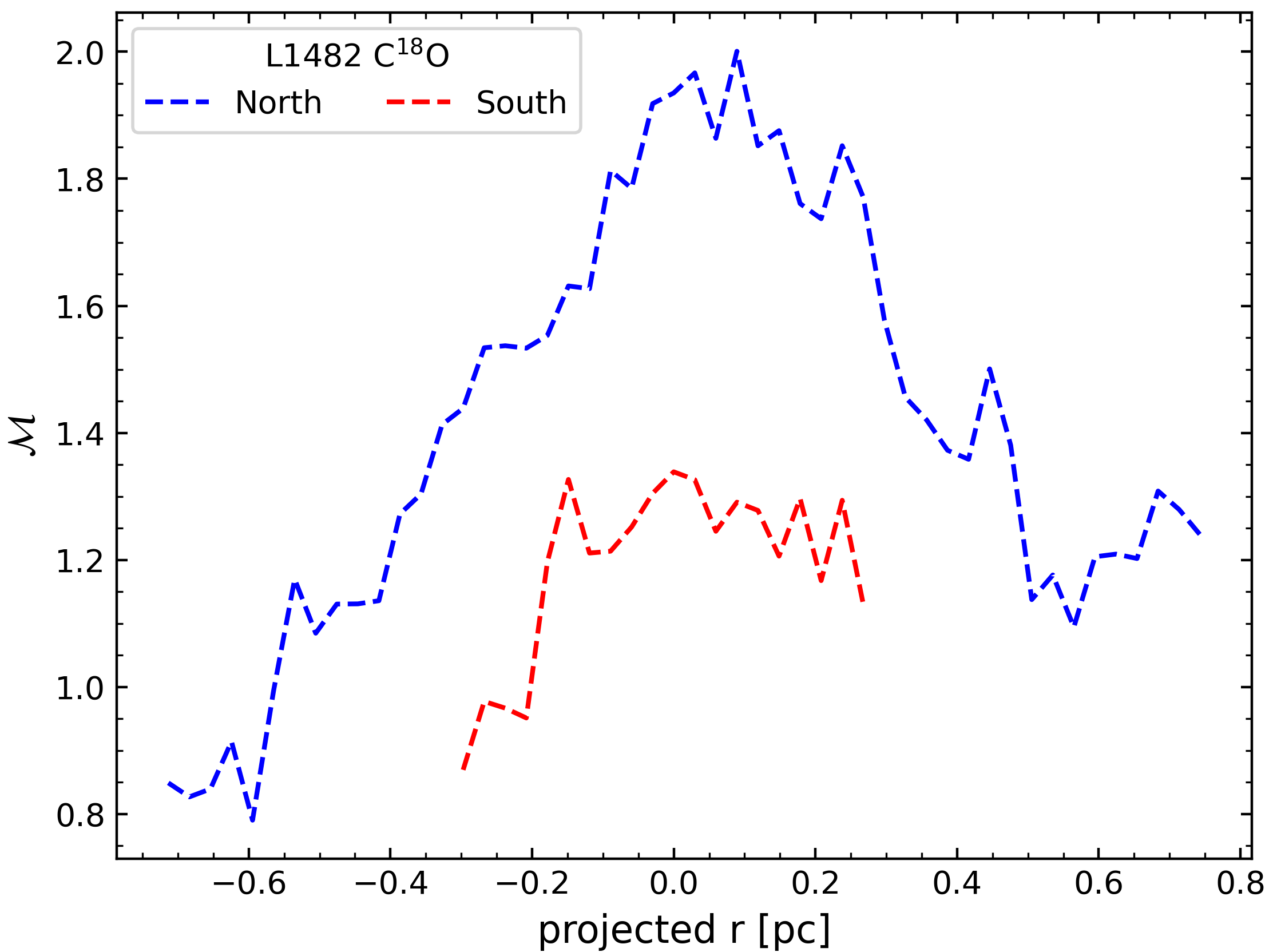}
    \caption{\co North (blue) and South (red)  Mach number ($\cal M$)
      profiles. The tracers present supersonic line-width 
      profiles, and large differences between the two regions.}
    \label{fig:mach_comparison}
    \end{minipage}%
  \end{figure}
  
\subsection{Mach number profiles across the filament}
\label{sub:mach}

Here we analyze the gas line-widths. Variations in the velocity
dispersion across the filament are useful for identifying variations
in the gas kinematics within the filament \citep[see,
e.g.,][]{federrath16}.  In the top right panel of
Figure~\ref{fig:tracers_moments} we show the \co second-moment map. In
what follows, we compare the line-width profiles in the North and
South regions of L1482.

\begin{figure*}
    \begin{minipage}{\textwidth}
    \centering
    \includegraphics[width=\textwidth]{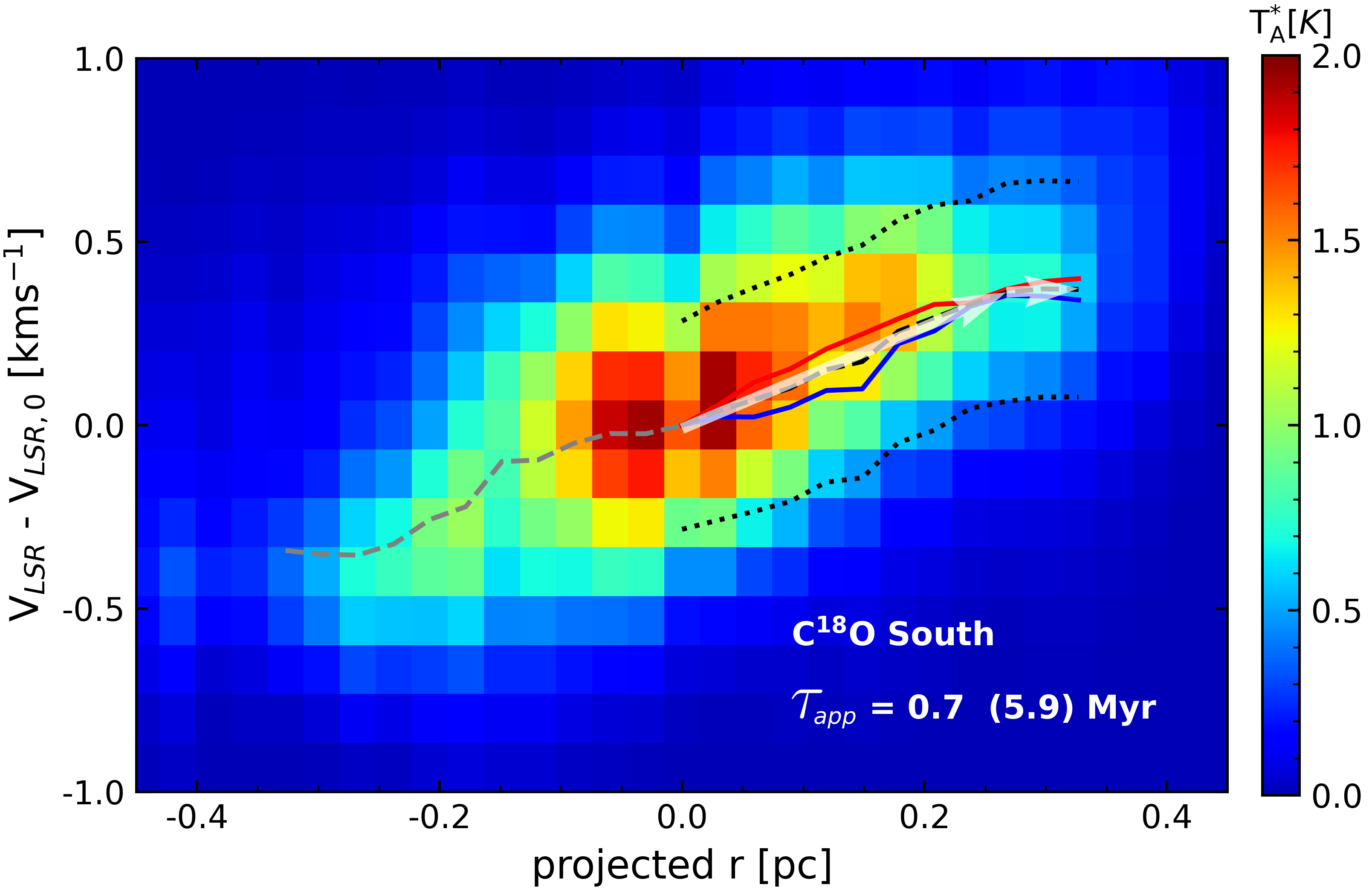}
    \caption{Velocity vs.\ projected radius map for \co L1482
      South. Negative r values indicate the east side of the filament
      while positive r values indicate the west side, relative to the
      dust ridgeline. The dashed grey curve shows the velocity
      gradient at r~$\leq$~0~\pc, while the blue curve shows the same
      r~$\leq$~0~pc flipped to the r~$\geq$~0~pc side for comparison
      with the r~$\geq$~0~pc red-shifted velocity gradient (red curve)
      of the filament.  The dashed black curve shows the mean of the
      blue and red velocity gradients.  The dotted black curves show
      the mean upper and lower bounds on the radial velocity between
      the each side of the filament, as measured by the
      intensity-weighted standard deviation of the velocity. The
      translucent white arrows represent the velocity profile derived
      from the best-fit line to the black dashed curve (the mean
      observed velocity profiles). The velocity profile flattens at
      $r\,\sim\,0.25$~\pc, as captured by the slopes of the inner and
      outer arrows. The associated timescales ($\tau_{app}$~=
      1/VG$_{app}$, where VG$_{app}$ is the velocity gradient) are
      shown in the bottom right corner. }
    \label{fig:slopes}
    \end{minipage}
\end{figure*}

We measure non-thermal motions in the gas through the non-thermal line
width ($\sigma_{\rm{NT}}$) and Mach number ($\cal M$). We derive the
non-thermal line width as
$\sigma_{\rm{NT}}\,=\, \sqrt{\sigma_{obs}^2 - kT_{k}/m}$
\citep[][]{liu19,gonzalez19}, where $ \sigma_{obs}$ is obtained from
the moment 2 velocity dispersion data, $m$ is the mass of the
molecule, $k$ is Boltzmann constant, and $T_k$ is the gas kinetic
temperature.  We assume that T$_k$ is equal to the dust temperature
T$_{\rm{d}}$ in the filament, where the hydrogen densities are higher
than 10$^4$~cm$^{-3}$ and the gas and dust are coupled
\citep[e.g.,][]{lippok13}.  In appendix~\ref{appendix:b} we fit a
softened power-law profile to the \herschel $\td$ map (see
\S~\ref{sect:mass_analysis}).  We derive the Mach number profile
$\cal M$~=~$\sigma_{\rm{NT}}/\rm{c}_s$, where
c$_s\,=\,\sqrt{kT_k/\mu m_{\rm H}}$ is the sound speed, $\mu$~=~2.33
is the mean molecular weight, and $m_{\rm{H}}$ is the hydrogen atomic
mass \citep[][]{liu19,gonzalez19}.

In Figure~\ref{fig:mach_comparison} we present the mean \co Mach
number profiles as a function of distance from the \nh ridgeline in
the North and South regions separately \citep[see above and ][for a
similar analysis in the ISF]{gonzalez19}.  \co has a mildly supersonic
profile (${\cal M} \sim 1.2$) in the South but more elevated
supersonic profile in the North.  For \co North, the portion of the
filament with actual star formation, the profile peaks toward the
filament ridgeline, as opposed to decreasing with density toward the
center.  This centrally increasing trend is the opposite of that found
in the simulations of \citet[][]{federrath16}.  They measure the Mach
number profiles in simulations that aim to capture the primary agents
that will affect the velocities, such as gravity, turbulence, magnetic
fields, and jet and outflow feedback, all of which should be in
operation in the CMC/L1482 North filament.  In contrast, when we look
in the South, the Mach number profile is almost flat over the extent
that we are able to probe.  In either the North or the South, over the
spatial scales that we probe, we observe no transition to a subsonic
regime as reported in \citet[][]{federrath16}.  The elevated values of
$\cal{M}$ that we observe could be caused by various observational
effects, of which line-of-sight averaging and spatial resolution may
be the first-order culprits.  Both these effects may broaden the
measured line-widths to some extent.  However, with a higher density
tracer such as \co, we do not expect these effects to be dominant.

As already noted above and by \citet{gonzalez19}, our measured Mach
number profiles can be compared to simulated profiles, similar to those
presented in \citet[][]{federrath16} and in particular their Figure~5,
so long as the simulations capture the high mass regime (which the
\Citealt{federrath16} simulations do not).  Moreover, these should be
compared to observations in other filaments at the same mass regime,
such as those presented in \citet[][]{gonzalez19}. See discussion
below.

\section{\co Velocity gradient across the filament}
\label{sub:vel_grad}  

Here we characterize the velocity gradient of \co across the filament
in the South region, observed as a prominent feature in
Figure~\ref{fig:aligned_tracers}. Given the confined symmetry of the
gradient on either side of the ridgeline (see also
\ref{fig:tracers_moments}), the fact that it is detected along 
the entire $\delta$ range that we probe, and the lack of clear infall
detections on these scales (see above), this signature is consistent
with a rotational origin as opposed to arising from, for example,
cloud-cloud collisions \citep[e.g., ][]{olmi02,fukui18}.

We characterize the magnitude and spatial extent of this gradient as
follows. In Figure~\ref{fig:slopes} we show the \co velocity vs.\
radius map and corresponding gradients.  In order to measure the gradients
we must first construct mean aligned velocity vs.\ radius maps.  We
accomplish this by:
\begin{enumerate}
\item[1]{Aligning the data cube w.r.t. the
    \nh ridgeline in $\delta$ (see Figure~\ref{fig:aligned_tracers}).}
  \item[2]{Marginalizing over the aligned $\delta$ coordinates in the South
      region.}
    \item[3]{Fitting the gradient in projected radius vs.\
velocity.}
\end{enumerate}
We obtain the intensity-weighted mean velocity as a function of
radius, shown in Figure~\ref{fig:slopes} as the blue and red curves,
and the intensity-weighted standard deviation of the velocity distribution 
at each radii, represented with dotted black curves in the figure. Since 
we wish to compare a single velocity gradient, we average the blue-shifted 
(east side) and red-shifted (west side) velocity profiles (mean profiles 
are shown as the dashed black curve in Figure~\ref{fig:slopes}).  We then 
fit a line to these mean V(r) curves (shown as white arrows) to obtain the
velocity gradients (VG$_{app}$), or equivalently timescales
$\tau _{app} = 1/\rm{VG}_{app}$, presented as translucent white arrows
in the figure with corresponding inner (outer) timescales of
0.7~(5.9)~Myr.  As above, we must correct for the (unknown)
inclination of the filament by dividing VG$_{app}$ by the factor
$\cos(\theta)$, where $\theta$ is (as defined above) the inclination
of the filament relative to the plane of the sky.

Figure~\ref{fig:slopes} reveals that the \co South velocity profile
(and therefore associated gradients) is approximately anti-symmetric
about the \nh ridgeline (which defines $r=0$), as illustrated by the
similarity between the red- and flipped blue-shifted curves in the
diagram.  Moreover, the transition to the shallower gradient on at
$r\sim\,0.25~\pc$ is also anti-symmetric.  This high degree of
anti-symmetry lends very strong support to the rotational
interpretation of the velocity profile.  Moreover, the inner gradient
is approximately constant with radius.  This may imply, to first
order, solid-body rotation of the filament about the long axis.  We
return to this in \S~\ref{sect:discussion} below.  To second order,
the diagram exhibits departures from this simple model, which we
discuss below. In the North region, the \co velocity field has a more
complex structure (see Figure~\ref{fig:tracers_moments}).  Given that
the North region has more fragmentation and correspondingly a larger
number of protostars, and its M/L profile as a function of radius is
both larger in amplitude and shallower in slope, we interpret these
differences as due to the combined action of gravity and rotation, as
opposed to the South wherein rotation appears primary in setting the
transverse filament radial velocity field. We also note that some of
the variations between the North and South may arise from differing
but unknown average inclinations between the two regions.

In addition to the differences between the North and South, we find
that in particular, in the \co South diagram (Figure~\ref{fig:slopes})
the gradient exhibits a clear and regular radial dependence, mentioned
above.  In the inner portion of the filament the gradient appears
somewhat steeper, indicating a shorter timescale
($\tau_{app}~=~0.7$~Myr), while the outer portion transitions to a
shallower gradient and correspondingly longer timescales. At a radius
of $r\,\sim\,0.25$~\pc, the velocity profile flattens significantly,
departing from the regular velocity pattern that we observe at smaller
$r$.  This outer flatted region has a gradient consistent with
timescales of $\tau_{app}\,\sim5.9$~Myr (see discussion below).

Moreover, such structures will benefit from detailed modeling of the velocity field.
We defer this investigation to future work using, e.g., the POLARIS
line radiative transport modeling capabilities \citep[see
e.g.,][]{reissl18a,reissl18b}.

\section{Discussion}
\label{sect:discussion}

In Figure~\ref{fig:slopes} we have show that the filament exhibits a
clear, regular, and linear velocity profile pattern consistent with
rotation.  To first order, the inner portion of the velocity profile
is consistent with solid body rotation.  However, clear departures
from this simple model are obvious: the velocity profile has a clear
break at $r\,\sim\,0.25$~\pc, where the velocity profile transitions
to a shallower gradient.  Moreover, in
Figure~\ref{fig:mach_comparison} we show that the gas line-widths and
thus non-thermal motions are only moderately supersonic (with Mach
numbers $\sim$\,1.2).

The question that we must address now is how the gas motions, inferred
from the rotation signature (\S~\ref{sub:vel_grad}), compare to
gravity (\S~\ref{sect:mass_analysis}). We do this by comparing the
gravitational force to the centripetal force taking advantage of the
simplicity of solid-body rotation implied by the linear appearance (to
first order) of the gradients presented in Figure~\ref{fig:slopes}
(also see \S~\ref{sub:vel_grad}).  We estimate the centripetal force
as:
\begin{eqnarray}
    \textrm{F}_c(r) = m a_c;
    \qquad a_c = \frac{v^2}{r\cos^2(\theta)};
    \label{eq:fc}
\end{eqnarray}
where $m$ is the gas mass, $\theta$ is the inclination of the filament
relative to the plane-of-the-sky, and $v$ is the velocity profile at
each radii (see Figure~\ref{fig:slopes}). The force of gravity is
given by:
\begin{eqnarray} 
    \textrm{F}_g(r) =  m g(r) = m g_{app}(r) \cos(\theta),
    \label{eq:fg}
\end{eqnarray}
where $g_{app}(r) = \xi \left(\frac{r}{\pc}\right)^{\gamma-1}$ is the
gravitational acceleration from Equation~(\ref{grav_acc}) and
Table~\ref{table:mass_param}.  From Equations~(\ref{eq:fc}) and
(\ref{eq:fg}), we obtain the ratio
\begin{equation} \label{eq:ratio_forces}
\frac{{\rm F}_c}{{\rm F}_g} = \frac{v^2}{r g_{app}(r)\cos^3(\theta)}
= \frac{v^2\pc^{-1}}{\xi \cos^3(\theta)}\left(\frac{r}{\pc}\right)^{-\gamma},
\end{equation}
or
\begin{equation} \label{eq:ratio_forces2}
  \frac{{\rm F}_c}{{\rm F}_g} =\frac{\pc}{\tau_{app}^2  \xi \cos^3(\theta)}\left(\frac{r}{\pc}\right)^{2-\gamma},
 \end{equation}
where $\xi$ and $\gamma$ are listed in Table~\ref{table:mass_param}.

\begin{figure}
     \begin{minipage}{\columnwidth}
     \centering
     \includegraphics[width=8.5cm]{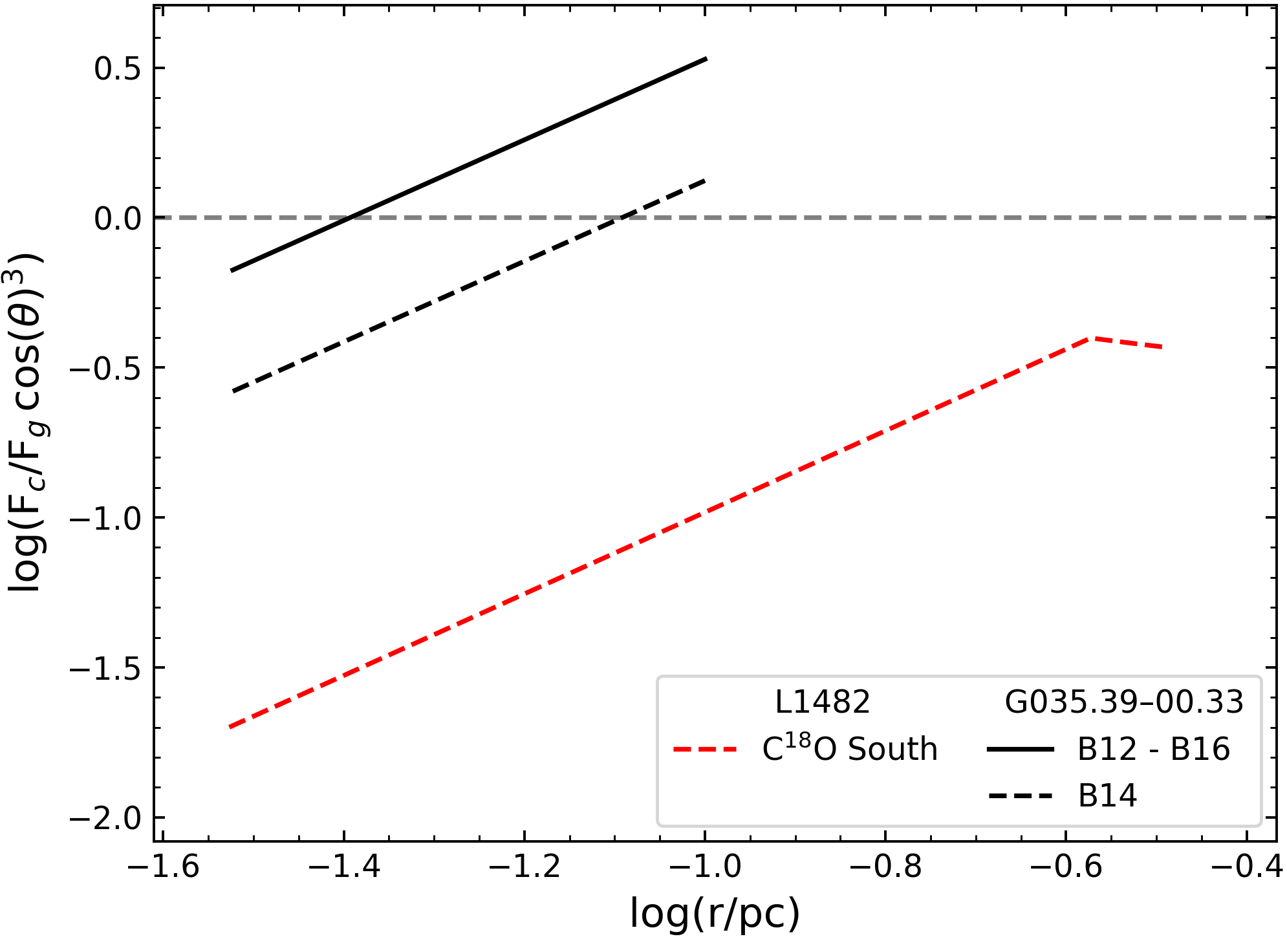}
     \caption{Ratio between the centripetal (F$_c$) and gravitational
       (F$_g$) forces (see Eq.~\ref{eq:ratio_forces}). We represent
       the \co South profile with a dashed red line. We include the
       profiles of slices B12-B16 (black solid line) and B14 (black
       dashed line) from the IRDC G035 \citep[][]{henshaw14}.  We show
       F$_c$/F$_g$\,=\,1 with a dashed gray line. These curves, taken
       at face value, assume $\cos(\theta)=1$, that is, that the
       filament is not inclined relative to the plane-of-the-sky (see
       text).  For \co South we see that gravity dominates over
       rotation, presenting a turning point at r~$\sim$~0.25~pc. If
       the inclination of the filament is significant, then rotation
       will be more dominant.}
    \label{fig:ratio_forces}
    \end{minipage}%
\end{figure}

In Figure~\ref{fig:ratio_forces} we present the apparent (plane of the
sky) $\rm{F}_c/\rm{F}_g$ profile for the South portion of the
filament as traced by \co.  The basic features of this diagram show
that the role of rotation relative to gravity increases with radius
until $r\,\sim\,0.25$~\pc, inside of which the timescales are
$\sim\,0.7$~Myr.  At $r\,\sim\,0.25$~\pc\ gravity takes over,
regulating the filament structure just when the rotational profile
would imply break-up or near break-up of the filament.

  Moreover, the velocity gradient (on both sides of the filament)
  is very smooth and regular (see Figure~\ref{fig:slopes}), which can 
  be interpreted as the filament being in a dynamically relaxed state.  
  To arrive at such a state, the filament must have undergone a $\sim$
  couple of orbits.  This then raises the question of timescales.
  The outer timescale ($\sim\,6$~Myr) is long compared to the inner
  one, implying that this structure is stable and long-lived, with a
  lifetime of $\sim$ a few times $6$~Myr.  \citet{gomez14} and
  \citet{gomez18}, based on simulations with and without magnetic
  fields, find stable and long-lived filamentary structures with
  approximately similar lifetimes. They attribute these longer
  timescales to the stability imparted by mass accretion from the
  immediate environment.  This agreement between the timescales we
  measure here and the previously simulated ones is encouraging.
  However, the comparison may be limited as the simulated filaments in
  \citet{gomez14} and \citet{gomez18} have lower line masses
  ($\sim\,40\,$\msun\pc$^{-1}$), are embedded in lower mass clouds, and
  do not show obvious rotation signatures.

If the inclination is non-zero, the role of rotation will be larger
compared to gravity, driving the outer force ratio profile closer to
unity, and thus closer to break-up.  The basic appearance of this
diagram therefore implies outside-in evolution of this filamentary
structure as the action of gravity and the removal of angular momentum
progressively remove support, allowing for progressive mass
concentration (see Figure~\ref{fig:enclosed-line-mass}) and eventual
star formation to take place.

In Figure~\ref{fig:ratio_forces} we also include the
$\rm{F}_c/\rm{F}_g$ profile for the
G035 \citep[e.g.,][]{henshaw13,jimenez14} N$_2$H$^+$ data presented in
\citet{henshaw14}, which has been previously proposed to be in an
early evolutionary state.  The gas kinematics in this filament exhibit
a velocity gradient, with an east to west orientation, similar to
L1482. These authors measure a gradient of $-$13.9~$\kms$~pc$^{-1}$,
for which $\rm{F}_c/\rm{F}_g$ profile is presented as the solid black line
in Figure~\ref{fig:ratio_forces}, labeled as ``B12 - B16'' \citep[see
appendix of ][]{henshaw14}.  However, we note that their figures are
consistent with a somewhat shallower average gradient of
$-$8.7~$\kms$~pc$^{-1}$, which $\rm{F}_c/\rm{F}_g$ profile, which we
present as a dashed black line in the same figure.  Whatever measure
of rotation, G035 appears much more rotationally dominated compared to
the L1482 filament.  This may partially be caused by the higher
density tracer used by \citet{henshaw14}, or the proposed early
evolutionary state of the G035 filament \citep{henshaw13}.  If the latter, 
the G035 filament may not have had time to dissipate its angular momentum to
collapse toward cluster formation.  Another possibility is that the
inclinations of L1482 versus G035 are different, which may account for
some of the observed differences. At present we have no strong constraints 
on the inclination for either system.

If the L1482 filament inclination is significant, then rotation will
play a larger role.  In \S~\ref{sect:dist} we find no significant
correlation between $\delta$ and $\pi$ over the filament as a whole.
As discussed, this represents a weak constraint on the system
inclination due in part to the small number of \gaia-detected YSOs and
their corresponding $\pi$ errors.  Meanwhile, the clear ``zig-zag''
plane-of-the-sky morphology of the South region as a whole may
indicate a corkscrew or ``pig tail''-like morphology (and see below).
Such undulations would cause us to again underestimate the role of
rotation relative to that of gravity.  Hence the basic appearance of
Figure~\ref{fig:ratio_forces} is consistent with rotation playing a
significant role in the South while being comparatively less dominant
in the North, which is clearly gravitationally dominated since it is
presently forming stars, and has a larger accompanying line-mass
profile.

\begin{figure}
     \begin{minipage}{\columnwidth}
     \centering
     \includegraphics[width=8.5cm]{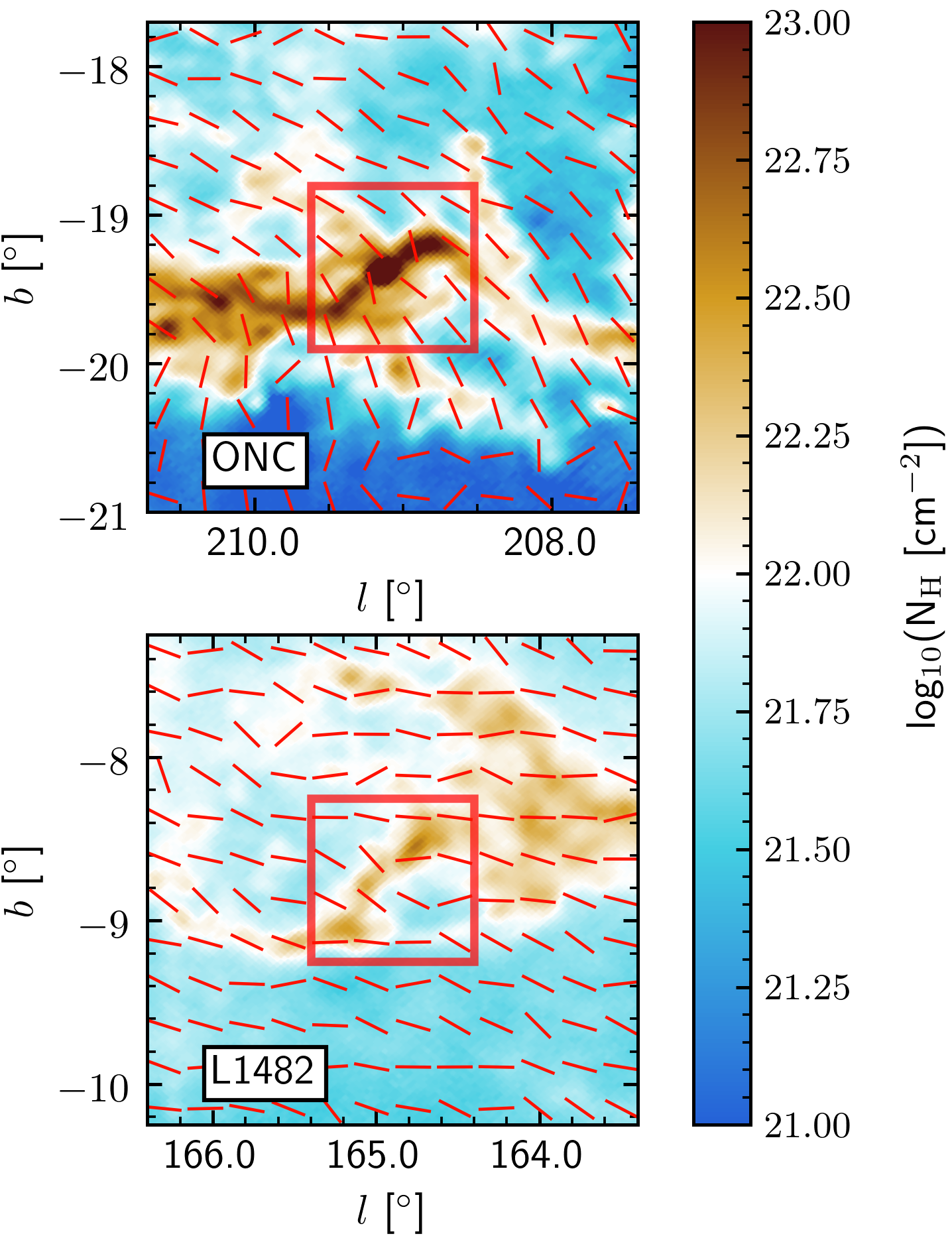}
     \caption{Planck \nh map (background) of the Orion ONC (top) and
       L1482 (bottom), both regions highlighted with red boxes. Red
       lines show the polarization vectors rotated by 90$^{\circ}$ to
       indicate the orientation of the magnetic field projected on the
       plane of the sky. For both the ONC and L1482, the projected
       magnetic field is perpendicular to the filaments.}
    \label{fig:planck}
    \end{minipage}%
\end{figure}

A velocity feature similar to that measured in L1482 has been studied
in the Orion~A ISF, reported by \citet[][]{gonzalez19} (and references
therein) which they interpret as rotation.  Based primarily on $^{12}$CO
position-velocity (PV) diagrams, they identify two velocity components
in the northern half of the ISF (see their Figure~8 and \S~4.4) with
estimated mean spatial separations of $\sim 1.3$~pc and angular
velocities (assuming circular rotation) of $\sim 1.4$~Myr$^{-1}$.  On
smaller scales, they also present \nhp ``bumps and wiggles'' (see
their Figure~4) that appear consistent with torsional-type structures
with short timescales.  While the relation between the larger $^{12}$CO
and smaller scale \nhp velocity features in Orion~A is not presently
understood, they both present the appearance of rotational structures.
The features that we characterize here in L1482, in contrast, are
found in a lower density filament (see
Figure~\ref{fig:enclosed-line-mass}), are closer to the central axis
of the filament (the \nh ridgeline) than the $^{12}$CO Orion velocity
feature, and may present different inner-filament timescales.

In terms of the magnetic field as probed by linear polarization, we
can compare L1482, Orion, and G035.  In G035 the projected magnetic
field orientation presented by \citet[][]{liu18} in the central
portion ``M'' (high density) is perpendicular to the filament. The
north and south portions (lower density) show a magnetic field almost
parallel to the filament. This trend has been studied in multiple
systems, including L1482 and the Orion~A Integral Shaped Filament
(ISF) region, containing the ONC. In Figure~\ref{fig:planck} we show
the Planck polarization data \citep[][]{planck13}, which exhibits a
projected field orientation that is preferentially perpendicular to
the main filament axis
\citep[e.g.,][]{soler19,planck16,li15,tahani2018}. 
The magnetic field morphology can be interpreted in 3D as either
helical or bow-like 
\citep[e.g.,][]{inoue2018,li_klein2019,tahani2018,tahani2019, gomez18}. Moreover, close inspection of Figure~\ref{fig:aligned_tracers} reveals that the northern portion of
L1482-South has more prominent blue-shifted velocities, while the
southern portion has more prominent red-shifted velocities, a
signature that may be consistent with a helical velocity field in a
filament with a 3D morphology similar to a corkscrew.  One key
question here is if the filament rotational motions may be linked to
the 3D structure of the magnetic field in these systems, as seen
  for example in L1641 \citep[][]{uchida91}.  The combination of
rotational motion about the long axis and a helical, or possibly
  bow--like (but see below), magnetic field morphology may be the
most natural explanation that accounts for the combination of the
observed gas radial velocities, the linear polarization patterns, and
the overall ``zig-zag'' morphology of the filament.  As discussed
above, we suggest that future modeling of molecular line emission
under various assumptions for the 3D density, velocity, and magnetic
field properties would shed light on the structures we observe
\citep[e.g.,][]{reissl18a,reissl18b} and potentially yield constraints
on the pitch-angle of the magnetic field in this system.

In terms of simulations, we find none, to the best of our knowledge,
that reproduce rotational signatures similar to the ones we observe.
In simulated turbulence and gravity-driven filament formation
\citep[e.g.,][]{priestly2020}, even when including 
magnetic fields \citep[e.g.,][]{gomez18,federrath16, li_klein2019,
  koertgen2017, seifried20,wareing20}, rotation is not
present. Moreover, these simulations probe low mass- and
mass-per-unit-length scales compared to $\sim 10^5$~$\msun$ clouds and
their correspondingly higher M/L filaments like California and Orion
(see Table~\ref{table:mass_param}).  See also \citet[][]{reissl20} for
a direct comparison between the line mass profiles in Orion and
California/L1482 compared to one example of an MHD simulation of a
filament.

  Moreover, \citet{inoue2018} and \citet{li_klein2019} simulate
  somewhat higher mass clouds or regions, including models for
  magnetic fields. \citet{li_klein2019} illustrate how the bow-shaped
  B-field morphology, in particular the directional flip on either
  side of the filament, is also consistent with the signature of a
  helical field along specific sight-lines (see their Figure~16).
  While they do not analyze the kinematics of the dense gas in detail,
  required for comparison to the specific signature presented here,
  they do detect a simulated flip in the velocity field on either side
  of the filament along certain sight-lines.  Furthermore, their
  simulated filament is mostly very straight along its extent,
  exhibiting only slight, gentle curvature and not the ``zig-zag''
  morphology that we observe here.  \citet{inoue2018} simulate the
  collapse in a small ($r = 1.5$~pc) but massive ($M = 838$~\msun)
  region.  Their simulation produces a more massive sink particle
  ($M \gtrsim 50$~\msun) on a extremely short timescale of
  $\sim\,0.45$~Myr.  Taking cuts across the sink particle, they
  recover some simulated velocity gradient; however, we observe no such
  analogies to massive sink particles (presumably these would be
  massive and very compact ``clumps'').  On the contrary, we have
  shown that our observations are consistent with a relatively smooth
  mass distribution along L1482-South.  In summary, the detection in
  simulations of features resembling velocity gradients are an
  intriguing possibility for explaining the basic velocity flip that
  we detect; however we must not only explain this flip, but also the
  magnitude compared to gravity, the outer break and transition to
  shallower gradients, and the basic filament properties.
  Furthermore,  these models require viewing the simulated filaments at 
  specific angles in order to be consistent with the data and reproduce
  flips about the ridgelines, which may be reasonable for randomly 
  oriented sets of filaments. However, the Gaia data (although this 
  is not a strong constraint at present) would seem to imply that 
  overall L1482, on larger scales, is not strongly inclined.

Meanwhile, lower density cosmological simulations also produce
non-rotating filaments that arise due to the action of gravity
\citep[e.g.][]{springel18}.  This is consistent with the idea that
gravo-turbulent filaments simply do not rotate.  In terms of
observations, the situation is less clear but the existing data point
toward a scenario where rotation appears not to be found in filaments
in lower mass clouds
  \citep[e.g.,][]{hacar16,hacar13,kirk13,shimajiri19} but was analyzed
  for example in the high mass Orion~A L1641 filament
  \citep{uchida91}.   In particular, in \citet{shimajiri19}, they
  study the velocity gradients in the Taurus B211/B213 in $^{12}$CO
  and $^{13}$CO (1-0). These authors interpret their velocity data
  with a model composed of two intersecting lower density sheets,
  observed at a specific viewing angle, and conclude that this
  particular system is well-reproduced by this configuration.  They
  also note that because their tracers are optically thick, they
  require optically thin dense gas tracers to probe the velocity field
  inside the filament.  As noted above, this is a straight filament,
  it is a lower mass and lower line-mass system: the Taurus B213 has
  an line mass of $\lambda_{app} \sim 100$~\msun/pc at a projected
  radius of $r \sim 1$~pc, about a factor of 2 lower than our measured
  $\lambda_{app} = 205$~\msun/pc measured at the same projected radius
  of $r = 1$~pc (see Table~\ref{table:mass_param}).  The morphologies of the two filaments
  are very dissimilar, one being straight and mildly curved, while the
  L1482-South filament has a plane-of-the-sky ``zig-zag''
  morphology. 

Gradients have also been observed and quantified in other low mass
  regions, such as Serpens-Main \citep[e.g.][]{olmi02, lee14}, using
  optically thin and high density gas tracers.  In particular,
  \citet{lee14} observe a perpendicular gradient in the ``FC1''
  filament in N$_2$H$^+$, with an apparently sharp velocity jump.
  However, their main focus was the gradient along the filaments, and
  they do not suggest a physical origin for the perpendicular velocity
  gradients.  In Serpens-South, a cloud assumed to be associated with
  Serpens-Main, \citet{fernandez14} find perpendicular gradients in
  N$_2$H$^+$ for filament ``FE'' (11.9 km s$^{-1}$ pc$^{-1}$) and
  ``NFW'', but also dot not compare the velocity gradients to gravity.
  They postulate that the origin of these gradients is accretion from
  the local filament environment, and a result of projection in a flat
  structure that is inclined.  In order to quantify how these gradients 
  and their timescales compare with gravity and the L1482 filament, we 
  suggest they should be revisited in combination with M/L profiles, 
  similarly to the method we present in this paper. We also note that 
  the simulations outlined above, and the low line-mass filaments 
  analyzed in these observational studies, do not exhibit the 
  two-dimensional ``zig-zag'' morphology that we observe in L1482.
  However, see \citet{wareing20} (in particular their Figure~6) for 
  simulated filaments with "wiggles" which may be comparable to the 
  "zig-zag" structure we observe.  However, the \citet{wareing20} 
  numerical experiments do not reveal rotation, emphasizing the 
  importance of a comparison to all observational constraints 
  presented here, and in particular the kinematic signatures 
  presented in Figure~\ref{fig:slopes}.

Moreover, the lack of alignment between YSO outflows and filament
orientation reported in Perseus \citep[][]{stephens17} may be
consistent with the idea that lower mass clouds have more elevated
levels of turbulent (and thus randomized velocity) support compared to
higher mass clouds, as suggested by \citet[][]{stutz16}. In contrast
to Perseus, \citet[][]{kong19} report the detection of
filament-outflow orientation anti-correlation in the massive (M
$\sim10^5~\msun$) G28.37$+$0.07 cloud.  This may support the view
presented in \citet[][]{stutz16} that the internal evolution of high
mass clouds may be qualitatively different than low mass ones,
specifically in relation to the role of rotation and the magnetic
field in the collapse process.

Given the lack of rotation in lower mass turbulence simulations and
lower mass clouds, the observations we present here would then lead to
the suggestion that there must be some other mechanism for filament
rotation in higher mass clouds.  As mentioned above, the magnetic
field is the only agent capable of playing this role \citep[see e.g.,
][and references therein]{hennebelle18,schleicher2018}. It may
potentially act via magnetic breaking (by analogy to stars).  This
would imply that all filaments that are rotating must remove inner
angular momentum via the magnetic field in order to collapse.  The
magnetic field would transport this rotational energy from small
scales near the filament spines to larger radii.  The observation of
the more evolved L1482-North filament within this same cloud and just
to the north, with fragmentation, accompanying protostar formation and
no clear rotational signature, may support this idea of filament
evolution.  On the other hand, if the rotational signature is an
imprint set by the earlier and diffuse ISM phase of cloud evolution,
the question remains as to why lower mass clouds and their lower
line-mass but still star-forming filaments are not rotating.  Either
the filaments where never rotating, which could be related to their
mass-scale as suggested here, or observations and simulations should
be able to capture an earlier lower-density phase of angular momentum
dissipation.

We speculate that the structures reminiscent of helical or double
helix loops seen here (see Figure~\ref{pv}), in Orion~A
\citep[][]{gonzalez19} and possibly to be seen in the G28.37$+$0.07
cloud \citep[][]{kong19}, are the anchors of the magnetic fields.  In
this case, these would also be inevitable features of massive star
forming filaments.  If correct, this would make magnetic fields an
essential feature of (massive) star and cluster forming filaments.
Linking the rotational signatures to the magnetic field naturally
leads to a torsional, helical, or bow--like field morphology. 
The observed velocity flip perpendicular to the L1482-South ridgeline
(Figure~\ref{fig:aligned_tracers}) is also accompanied by a flip in
the magnetic field \citep[][]{tahani2018} in California on larger
scales, a flip that these authors interpreted as a possible signature
of magnetic field helicity. The connection between the
  large-scale inferred field geometry and the small scale velocity
  feature we detect is not presently understood and must be
  investigated in future work.

\section{Conclusions}
\label{sect:conc}

Previous studies have shown that the California molecular cloud is
similar to Orion~A in terms of mass (M $\sim 10^5$\,\msun),
morphology, undulating filaments, and magnetic field geometry but
strikingly different in YSO content
\citep[][]{lada09,megeath12,kong15,lada17,tahani2018}.  We present the
study of the filament L1482 located in California. L1482 is one of the
most massive filaments in California and contains $\sim$~56\% of the
YSOs in the cloud. Here we characterize the physical properties of
this filament by combining the analysis of YSO distance information,
gas mass distributions, and gas velocities. We
summarize our main results as follows.\\
\noindent$\bullet$ We find a distance of 511$^{+17}_{-16}$~pc based on
\gaia DR2 astrometry of YSOs in the filament.\\
\noindent$\bullet$ Using \herschel and \planck \nh maps, we
characterize the projected mass per unit length (M/L) of the L1482
filament as a whole, and divide the filament into two distinct
regions based on differences in the presence of star formation, M/L
profiles, morphology, and gas velocities (see below): the South, with
a lower M/L, no star formation, a plane-of-the-sky ``zig-zag''
morphology, and coherent velocity patterns, versus the North, with
star formation, a higher M/L profile, and more chaotic
velocity patterns.\\
\noindent$\bullet$ All the above regions of the filament have M/L
profiles consistent with a scale-free power-law shape.  At 1~pc, the
filament as a whole, the North, and the South regions have M/L
power-law normalizations of 214, 264, and 205\,$\msun\rm{pc}^{-1}$ and
power-law indices of 0.62, 0.51, and 0.64, respectively.  See
Table~\ref{table:mass_param} and Figure~\ref{fig:enclosed-line-mass} 
for full M/L profiles. \\
 \noindent$\bullet$ Based on the M/L profiles, and assuming
  cylindrical geometry and axial symmetry motivated by the mass
  distribution (see Figures~\ref{cumulative-mass} \& 
  \ref{fig:enclosed-line-mass}), we calculate the volume density,
  gravitational potential, and gravitational acceleration
  profiles. See
  Table~\ref{table:mass_param}.  \\
\noindent$\bullet$ We present intensity-weighted position-velocity
diagrams of IRAM~30~m observations of \co, \hco, HNC, and \nhp
(1-0). These reveal complex twisting and turning velocity structures,
with timescales of order $\sim\,1$~Myr.  These
undulations are particularly striking in the South of L1482 in \co.\\
 \noindent$\bullet$ We detect a velocity profile in \co consistent
  with rotation about the long-axis of the South filament.  The
  velocity profile pattern is regular and smooth.  It profile ``hugs''
  the filament, that is, it is confined to within
  $r \lesssim \pm 0.4$~pc from the \nh ridgeline and is anti-symmetric
  about the ridgeline.  See Figure~\ref{fig:slopes}.\\
\noindent$\bullet$ The above velocity profile grows linearly to
$r~\sim~0.25$~\pc, outside of which it softens to shallower slope.
The inner ($r~\leq~0.25$~\pc) timescale of the velocity gradient is 0.7~Myr.  The
outer ($r~>~0.25$~\pc) timescale is much longer, about 6~Myr.\\
\noindent$\bullet$ When compared to the gravitational field, and using
a simple solid body rotation toy model, the rotational profile implies
that the filament approaches break-up at $r\sim0.25$~\pc, where it
transitions to being gravity-regulated.  See
Figure~\ref{fig:ratio_forces}. \\
\noindent$\bullet$ The long timescales on the outside of the
rotational profile, combined with the break in the rotational profile,
imply that the filament structure is stable and long lived (few times
6~Myr) and that the filament is undergoing outside-in evolution in the
process of dissipating its angular momentum toward collapse and
eventual star formation (as is observed in the North portion of the
larger L1482
filamentary structure).  \\
  \noindent$\bullet$ The above results assume the filament is
  oriented in the plane of the sky.  While the filament inclination is
  not well-constrained, the role of rotation compared to gravity will
  only be greater if the inclination is non-zero.\\
   \noindent$\bullet$ The \planck polarization vectors, rotated to
  infer the plane-of-the-sky magnetic field geometry, are
  approximately perpendicular to the to main filament axis, as is also
  seen in other clouds such as Orion~A and in G035.  The appearance of
  the position-velocity diagrams, the rotational signature discussed
  above, and the plane-of-the-sky ``zig-zag'' morphology of the
  filament, taken together with \planck, may imply a 3D ``corkscrew''
  filament with a helical magnetic field morphology could be a natural
  explanation of the various observational constraints.  Meanwhile, a
  ``bow'' shaped field (as reproduced in simulations) may also account
  for certain observational features, such as the flip in the magnetic 
  field along the line-of-sight and the magnetic field in the POS being 
  perpendicular to the filament.
  
Comparison to turbulence and MHD simulations (which do not show rotation
  in filaments), observations of lower mass filaments (which also do not show 
  rotation, see the related discussion in \S~\ref{sect:discussion}), previous 
  results in Orion~A, constraints from e.g., the G035 filament, together lead 
  to the suggestion that high mass (M $\sim 10^5$\,\msun) clouds host
  rotating filaments that may be more common than previously thought.
  In this hypothesis, these structures must undergo angular momentum
  evolution (e.g., \Citealt{hennebelle18} and references therein), for example
  losing angular momentum through rotation, in order to collapse as they 
  progress toward star cluster formation.

We speculate that the observed velocity flip across the filament
  on small scales may be related to the flip in the magnetic field
  observed in California on far larger scales
  \citep[][]{tahani2018}. We emphasize that the connection between the
  large scale signatures of field geometry and the small scale
  velocity features detected here are not presently understood.
  Hence, further investigation into the scale dependence of the
  magnetic field strengths and geometries are essential.  Moreover,
  future observational work should include scrutiny of larger portions
  of California to more broadly characterize the gas kinematics.
  Wide-field observations that trace magnetic fields on filament
  scales (capturing scales of $\lesssim$0.1~pc to $\gtrsim$10~pc and
  cloud scales, including Zeeman when possible, are essential for
  obtaining further constraints on both field geometry and strength.
  Constraining gas kinematics in combination with constraints for the
  magnetic field is critical.  Scrutiny of the YSO and stellar
  kinematics, including radial velocities, and their possible link to
  the cloud kinematics, will explore the link between the stellar and
  gas content \citep[][]{stutz16}. Besides California and Orion~A, it
  is imperative to kinematically characterize other high-mass
  filaments like e.g., G28.37$+$0.07 \citep[][]{kong19}.  Finally,
  numerical work should focus on molecular line modeling of the
  velocity field of "zig-zag'' filaments to elucidate the 3D structure
  filaments that have a significant degree of rotation \citep[see
  e.g.,][]{reissl18a,reissl18b}.

\section*{Acknowledgments}

The authors thank the referee for helpful comments that improved the
text.  RA and AS thank Andrew P.\ Gould for very helpful discussions.
The authors thank Shuo Kong, Charlie Lada, and John Bieging for
generously sharing their SMT \sco data with us.  The authors thank
Jouni Kainulainen for kindly sharing the G035 8\mum extinction map.
In addition, we thank Valentina Gonz\'alez-Lobos, Paola Casseli,
  S. Thomas Megeath, Enrique Vázquez- Semadeni, and Doug Johnstone for
  helpful discussions. RA acknowledges funding support from CONICYT
Programa de Astronomía Fondo ALMA-CONICYT 2017 31170002. AS gratefully
acknowledges funding support through Fondecyt Regular (project code
1180350).  AS, NL, and RR acknowledge funding support from Chilean
Centro de Excelencia en Astrofísica y Tecnologías Afines (CATA) BASAL
grant AFB-170002. NL is gratefully supported by a Fondecyt Iniciacion
grant (11180005).  HLL acknowledges the funding from Fondecyt
Postdoctoral (project code 3190161). SR and RSK acknowledge
  financial support from the German Research Foundation (DFG) via the
  Collaborative Research Center (SFB 881, Project-ID 138713538) 'The
  Milky Way System' (subprojects A1, B1, B2, and B8).  They also thank
  for funding from the Heidelberg Cluster of Excellence STRUCTURES in
  the framework of Germany's Excellence Strategy (grant EXC-2181/1 -
  390900948) and for funding from the European Research Council via
  the ERC Synergy Grant ECOGAL (grant 855130). RR acknowledges
support from Conicyt-FONDECYT through grant 1181620. This work is
  based on observations carried out under projects number 031-18,
  152-18, 151-19 with the IRAM 30m telescope. IRAM is supported by
  INSU/CNRS (France), MPG (Germany) and IGN (Spain).

\bibliographystyle{apj}
\bibliography{ref}

\begin{thebibliography}{}
\expandafter\ifx\csname natexlab\endcsname\relax\def\natexlab#1{#1}\fi

\bibitem[{{Abreu-Vicente} {et~al.}(2017){Abreu-Vicente}, {Stutz}, {Henning},
  {Keto}, {Ballesteros-Paredes}, \& {Robitaille}}]{abreu17}
{Abreu-Vicente}, J., {Stutz}, A., {Henning}, T., {et~al.} 2017, \aap, 604, A65

\bibitem[{{Andrews} \& {Wolk}(2008)}]{andrews08}
{Andrews}, S.~M., \& {Wolk}, S.~J. 2008, {The LkH{\ensuremath{\alpha}} 101
  Cluster}, ed. B.~{Reipurth}, Vol.~4, 390

\bibitem[{{Arenou} {et~al.}(2018){Arenou}, {Luri}, {Babusiaux}, {Fabricius},
  {Helmi}, {Muraveva}, {Robin}, {Spoto}, {Vallenari}, {Antoja},
  {Cantat-Gaudin}, {Jordi}, {Leclerc}, {Reyl{\'e}}, {Romero-G{\'o}mez}, {Shih},
  {Soria}, {Barache}, {Bossini}, {Bragaglia}, {Breddels}, {Fabrizio},
  {Lambert}, {Marrese}, {Massari}, {Moitinho}, {Robichon}, {Ruiz-Dern},
  {Sordo}, {Veljanoski}, {Eyer}, {Jasniewicz}, {Pancino}, {Soubiran}, {Spagna},
  {Tanga}, {Turon}, \& {Zurbach}}]{arenou18}
{Arenou}, F., {Luri}, X., {Babusiaux}, C., {et~al.} 2018, \aap, 616, A17

\bibitem[{{Caselli} {et~al.}(2002){Caselli}, {Benson}, {Myers}, \&
  {Tafalla}}]{caselli2002}
{Caselli}, P., {Benson}, P.~J., {Myers}, P.~C., \& {Tafalla}, M. 2002, \apj,
  572, 238

\bibitem[{{Choi} {et~al.}(2020){Choi}, {Austermann}, {Basu}, {Battaglia},
  {Bertoldi}, {Chung}, {Cothard}, {Duff}, {Duell}, {Gallardo}, {Gao}, {Herter},
  {Hubmayr}, {Niemack}, {Nikola}, {Riechers}, {Rossi}, {Stacey}, {Stevens},
  {Vavagiakis}, {Vissers}, \& {Walker}}]{primecam}
{Choi}, S.~K., {Austermann}, J., {Basu}, K., {et~al.} 2020, Journal of Low
  Temperature Physics, 199, 1089

\bibitem[{{Contreras} {et~al.}(2013){Contreras}, {Rathborne}, \&
  {Garay}}]{contreras13}
{Contreras}, Y., {Rathborne}, J., \& {Garay}, G. 2013, \mnras, 433, 251

\bibitem[{{Dame} {et~al.}(2001){Dame}, {Hartmann}, \& {Thaddeus}}]{dame01}
{Dame}, T.~M., {Hartmann}, D., \& {Thaddeus}, P. 2001, \apj, 547, 792

\bibitem[{{Evans} {et~al.}(2015){Evans}, {Di Francesco}, {Lee}, {J{\o}rgensen},
  {Choi}, {Myers}, \& {Mardones}}]{evans15}
{Evans}, Neal~J., I., {Di Francesco}, J., {Lee}, J.-E., {et~al.} 2015, \apj,
  814, 22

\bibitem[{{Falgarone} {et~al.}(2001){Falgarone}, {Pety}, \&
  {Phillips}}]{falgarone01}
{Falgarone}, E., {Pety}, J., \& {Phillips}, T.~G. 2001, \apj, 555, 178

\bibitem[{{Federrath}(2016)}]{federrath16}
{Federrath}, C. 2016, \mnras, 457, 375

\bibitem[{{Fern{\'a}ndez-L{\'o}pez} {et~al.}(2014){Fern{\'a}ndez-L{\'o}pez},
  {Arce}, {Looney}, {Mundy}, {Storm}, {Teuben}, {Lee}, {Segura-Cox}, {Isella},
  {Tobin}, {Rosolowsky}, {Plunkett}, {Kwon}, {Kauffmann}, {Ostriker}, {Tassis},
  {Shirley}, \& {Pound}}]{fernandez14}
{Fern{\'a}ndez-L{\'o}pez}, M., {Arce}, H.~G., {Looney}, L., {et~al.} 2014,
  \apjl, 790, L19

\bibitem[{{Fiege} \& {Pudritz}(2000{\natexlab{a}})}]{fiege20a}
{Fiege}, J.~D., \& {Pudritz}, R.~E. 2000{\natexlab{a}}, \mnras, 311, 85

\bibitem[{{Fiege} \& {Pudritz}(2000{\natexlab{b}})}]{fiege20b}
---. 2000{\natexlab{b}}, \mnras, 311, 105

\bibitem[{{Fischer} {et~al.}(2017){Fischer}, {Megeath}, {Furlan}, {Ali},
  {Stutz}, {Tobin}, {Osorio}, {Stanke}, {Manoj}, {Poteet}, {Booker},
  {Hartmann}, {Wilson}, {Myers}, \& {Watson}}]{fischer17}
{Fischer}, W.~J., {Megeath}, S.~T., {Furlan}, E., {et~al.} 2017, \apj, 840, 69

\bibitem[{{Fukui} {et~al.}(2018){Fukui}, {Torii}, {Hattori}, {Nishimura},
  {Ohama}, {Shimajiri}, {Shima}, {Habe}, {Sano}, {Kohno}, {Yamamoto},
  {Tachihara}, \& {Onishi}}]{fukui18}
{Fukui}, Y., {Torii}, K., {Hattori}, Y., {et~al.} 2018, \apj, 859, 166

\bibitem[{{Gaia Collaboration} {et~al.}(2018{\natexlab{a}}){Gaia
  Collaboration}, {Babusiaux}, {van Leeuwen}, {Barstow}, {Jordi}, {Vallenari},
  {Bossini}, {Bressan}, {Cantat-Gaudin}, {van Leeuwen}, \& et~al.}]{gaia18}
{Gaia Collaboration}, {Babusiaux}, C., {van Leeuwen}, F., {et~al.}
  2018{\natexlab{a}}, \aap, 616, A10

\bibitem[{{Gaia Collaboration} {et~al.}(2018{\natexlab{b}}){Gaia
  Collaboration}, {Brown}, {Vallenari}, {Prusti}, {de Bruijne}, {Babusiaux},
  {Bailer-Jones}, {Biermann}, {Evans}, {Eyer}, \& et~al.}]{gaiadr2}
{Gaia Collaboration}, {Brown}, A.~G.~A., {Vallenari}, A., {et~al.}
  2018{\natexlab{b}}, \aap, 616, A1

\bibitem[{{Ginsburg} \& {Mirocha}(2011)}]{ginsburg2011}
{Ginsburg}, A., \& {Mirocha}, J. 2011, {PySpecKit: Python Spectroscopic
  Toolkit}, Astrophysics Source Code Library, ascl:1109.001

\bibitem[{{G{\'o}mez} \& {V{\'a}zquez-Semadeni}(2014)}]{gomez14}
{G{\'o}mez}, G.~C., \& {V{\'a}zquez-Semadeni}, E. 2014, \apj, 791, 124

\bibitem[{{G{\'o}mez} {et~al.}(2018){G{\'o}mez}, {V{\'a}zquez-Semadeni}, \&
  {Zamora-Avil{\'e}s}}]{gomez18}
{G{\'o}mez}, G.~C., {V{\'a}zquez-Semadeni}, E., \& {Zamora-Avil{\'e}s}, M.
  2018, \mnras, 480, 2939

\bibitem[{{Gonz{\'a}lez Lobos} \& {Stutz}(2019)}]{gonzalez19}
{Gonz{\'a}lez Lobos}, V., \& {Stutz}, A.~M. 2019, \mnras, 489, 4771

\bibitem[{{Gould}(2003)}]{gould04}
{Gould}, A. 2003, arXiv e-prints, astro

\bibitem[{{Griffin} {et~al.}(2010){Griffin}, {Abergel}, {Abreu}, {Ade},
  {Andr{\'e}}, {Augueres}, {Babbedge}, {Bae}, {Baillie}, {Baluteau}, {Barlow},
  {Bendo}, {Benielli}, {Bock}, {Bonhomme}, {Brisbin}, {Brockley-Blatt},
  {Caldwell}, {Cara}, {Castro-Rodriguez}, {Cerulli}, {Chanial}, {Chen},
  {Clark}, {Clements}, {Clerc}, {Coker}, {Communal}, {Conversi}, {Cox},
  {Crumb}, {Cunningham}, {Daly}, {Davis}, {de Antoni}, {Delderfield}, {Devin},
  {di Giorgio}, {Didschuns}, {Dohlen}, {Donati}, {Dowell}, {Dowell}, {Duband},
  {Dumaye}, {Emery}, {Ferlet}, {Ferrand}, {Fontignie}, {Fox}, {Franceschini},
  {Frerking}, {Fulton}, {Garcia}, {Gastaud}, {Gear}, {Glenn}, {Goizel},
  {Griffin}, {Grundy}, {Guest}, {Guillemet}, {Hargrave}, {Harwit}, {Hastings},
  {Hatziminaoglou}, {Herman}, {Hinde}, {Hristov}, {Huang}, {Imhof}, {Isaak},
  {Israelsson}, {Ivison}, {Jennings}, {Kiernan}, {King}, {Lange}, {Latter},
  {Laurent}, {Laurent}, {Leeks}, {Lellouch}, {Levenson}, {Li}, {Li},
  {Lilienthal}, {Lim}, {Liu}, {Lu}, {Madden}, {Mainetti}, {Marliani}, {McKay},
  {Mercier}, {Molinari}, {Morris}, {Moseley}, {Mulder}, {Mur}, {Naylor},
  {Nguyen}, {O'Halloran}, {Oliver}, {Olofsson}, {Olofsson}, {Orfei}, {Page},
  {Pain}, {Panuzzo}, {Papageorgiou}, {Parks}, {Parr-Burman}, {Pearce},
  {Pearson}, {P{\'e}rez-Fournon}, {Pinsard}, {Pisano}, {Podosek}, {Pohlen},
  {Polehampton}, {Pouliquen}, {Rigopoulou}, {Rizzo}, {Roseboom}, {Roussel},
  {Rowan-Robinson}, {Rownd}, {Saraceno}, {Sauvage}, {Savage}, {Savini},
  {Sawyer}, {Scharmberg}, {Schmitt}, {Schneider}, {Schulz}, {Schwartz},
  {Shafer}, {Shupe}, {Sibthorpe}, {Sidher}, {Smith}, {Smith}, {Smith},
  {Spencer}, {Stobie}, {Sudiwala}, {Sukhatme}, {Surace}, {Stevens}, {Swinyard},
  {Trichas}, {Tourette}, {Triou}, {Tseng}, {Tucker}, {Turner}, {Vaccari},
  {Valtchanov}, {Vigroux}, {Virique}, {Voellmer}, {Walker}, {Ward}, {Waskett},
  {Weilert}, {Wesson}, {White}, {Whitehouse}, {Wilson}, {Winter}, {Woodcraft},
  {Wright}, {Xu}, {Zavagno}, {Zemcov}, {Zhang}, \& {Zonca}}]{Griffin-spire}
{Griffin}, M.~J., {Abergel}, A., {Abreu}, A., {et~al.} 2010, \aap, 518, L3

\bibitem[{{Hacar} {et~al.}(2016){Hacar}, {Kainulainen}, {Tafalla}, {Beuther},
  \& {Alves}}]{hacar16}
{Hacar}, A., {Kainulainen}, J., {Tafalla}, M., {Beuther}, H., \& {Alves}, J.
  2016, \aap, 587, A97

\bibitem[{{Hacar} {et~al.}(2013){Hacar}, {Tafalla}, {Kauffmann}, \&
  {Kov{\'a}cs}}]{hacar13}
{Hacar}, A., {Tafalla}, M., {Kauffmann}, J., \& {Kov{\'a}cs}, A. 2013, \aap,
  554, A55

\bibitem[{{Harvey} {et~al.}(2013){Harvey}, {Fallscheer}, {Ginsburg}, {Terebey},
  {Andr{\'e}}, {Bourke}, {Di Francesco}, {K{\"o}nyves}, {Matthews}, \&
  {Peterson}}]{calidr}
{Harvey}, P.~M., {Fallscheer}, C., {Ginsburg}, A., {et~al.} 2013, \apj, 764,
  133

\bibitem[{{Heiles}(1987)}]{heiles1987}
{Heiles}, C. 1987, {Interstellar Magnetic Fields}, ed. D.~J. {Hollenbach} \&
  J.~{Thronson}, Harley~A., Vol. 134, 171

\bibitem[{{Heiles}(1997)}]{heiles1997}
---. 1997, \apjs, 111, 245

\bibitem[{{Hennebelle}(2003)}]{hennebelle2003}
{Hennebelle}, P. 2003, \aap, 397, 381

\bibitem[{{Hennebelle}(2018)}]{hennebelle18}
---. 2018, Astrophysics and Space Science Library, Vol. 424, {Numerical
  Simulations of Cluster Formation}, ed. S.~{Stahler}, 39

\bibitem[{{Henshaw} {et~al.}(2014){Henshaw}, {Caselli}, {Fontani},
  {Jim{\'e}nez-Serra}, \& {Tan}}]{henshaw14}
{Henshaw}, J.~D., {Caselli}, P., {Fontani}, F., {Jim{\'e}nez-Serra}, I., \&
  {Tan}, J.~C. 2014, \mnras, 440, 2860

\bibitem[{{Henshaw} {et~al.}(2013){Henshaw}, {Caselli}, {Fontani},
  {Jim{\'e}nez-Serra}, {Tan}, \& {Hernandez}}]{henshaw13}
{Henshaw}, J.~D., {Caselli}, P., {Fontani}, F., {et~al.} 2013, \mnras, 428,
  3425

\bibitem[{{Herbig} {et~al.}(2004){Herbig}, {Andrews}, \& {Dahm}}]{herbig04}
{Herbig}, G.~H., {Andrews}, S.~M., \& {Dahm}, S.~E. 2004, \aj, 128, 1233

\bibitem[{{Inoue} {et~al.}(2018){Inoue}, {Hennebelle}, {Fukui}, {Matsumoto},
  {Iwasaki}, \& {Inutsuka}}]{inoue2018}
{Inoue}, T., {Hennebelle}, P., {Fukui}, Y., {et~al.} 2018, \pasj, 70, S53

\bibitem[{{Jim{\'e}nez-Serra} {et~al.}(2014){Jim{\'e}nez-Serra}, {Caselli},
  {Fontani}, {Tan}, {Henshaw}, {Kainulainen}, \& {Hernand ez}}]{jimenez14}
{Jim{\'e}nez-Serra}, I., {Caselli}, P., {Fontani}, F., {et~al.} 2014, \mnras,
  439, 1996

\bibitem[{{Kainulainen} \& {Tan}(2013)}]{kainulainen13a}
{Kainulainen}, J., \& {Tan}, J.~C. 2013, \aap, 549, A53

\bibitem[{{Kirk} {et~al.}(2013){Kirk}, {Myers}, {Bourke}, {Gutermuth},
  {Hedden}, \& {Wilson}}]{kirk13}
{Kirk}, H., {Myers}, P.~C., {Bourke}, T.~L., {et~al.} 2013, \apj, 766, 115

\bibitem[{{Kong} {et~al.}(2019){Kong}, {Arce}, {Maureira}, {Caselli}, {Tan}, \&
  {Fontani}}]{kong19}
{Kong}, S., {Arce}, H.~G., {Maureira}, M.~J., {et~al.} 2019, \apj, 874, 104

\bibitem[{{Kong} {et~al.}(2015){Kong}, {Lada}, {Lada},
  {Rom{\'a}n-Z{\'u}{\~n}iga}, {Bieging}, {Lombardi}, {Forbrich}, \&
  {Alves}}]{kong15}
{Kong}, S., {Lada}, C.~J., {Lada}, E.~A., {et~al.} 2015, \apj, 805, 58

\bibitem[{{K{\"o}rtgen} {et~al.}(2017){K{\"o}rtgen}, {Federrath}, \&
  {Banerjee}}]{koertgen2017}
{K{\"o}rtgen}, B., {Federrath}, C., \& {Banerjee}, R. 2017, \mnras, 472, 2496

\bibitem[{{Lada} {et~al.}(2017){Lada}, {Lewis}, {Lombardi}, \&
  {Alves}}]{lada17}
{Lada}, C.~J., {Lewis}, J.~A., {Lombardi}, M., \& {Alves}, J. 2017, \aap, 606,
  A100

\bibitem[{{Lada} {et~al.}(2009){Lada}, {Lombardi}, \& {Alves}}]{lada09}
{Lada}, C.~J., {Lombardi}, M., \& {Alves}, J.~F. 2009, \apj, 703, 52

\bibitem[{{Lada} {et~al.}(2010){Lada}, {Lombardi}, \& {Alves}}]{lada10}
---. 2010, \apj, 724, 687

\bibitem[{{Launhardt} {et~al.}(2013){Launhardt}, {Stutz}, {Schmiedeke},
  {Henning}, {Krause}, {Balog}, {Beuther}, {Birkmann}, {Hennemann},
  {Kainulainen}, {Khanzadyan}, {Linz}, {Lippok}, {Nielbock}, {Pitann}, {Ragan},
  {Risacher}, {Schmalzl}, {Shirley}, {Stecklum}, {Steinacker}, \&
  {Tackenberg}}]{laun13}
{Launhardt}, R., {Stutz}, A.~M., {Schmiedeke}, A., {et~al.} 2013, \aap, 551,
  A98

\bibitem[{{Law} {et~al.}(2020){Law}, {Li}, {Cao}, \& {Ng}}]{law20}
{Law}, C.~Y., {Li}, H.~B., {Cao}, Z., \& {Ng}, C.~Y. 2020, \mnras, 498, 850

\bibitem[{{Law} {et~al.}(2019){Law}, {Li}, \& {Leung}}]{law19}
{Law}, C.~Y., {Li}, H.~B., \& {Leung}, P.~K. 2019, \mnras, 484, 3604

\bibitem[{{Lee} {et~al.}(2014){Lee}, {Fern{\'a}ndez-L{\'o}pez}, {Storm},
  {Looney}, {Mundy}, {Segura-Cox}, {Teuben}, {Rosolowsky}, {Arce}, {Ostriker},
  {Shirley}, {Kwon}, {Kauffmann}, {Tobin}, {Plunkett}, {Pound}, {Salter},
  {Volgenau}, {Chen}, {Tassis}, {Isella}, {Crutcher}, {Gammie}, \&
  {Testi}}]{lee14}
{Lee}, K.~I., {Fern{\'a}ndez-L{\'o}pez}, M., {Storm}, S., {et~al.} 2014, \apj,
  797, 76

\bibitem[{{Li} {et~al.}(2014){Li}, {Esimbek}, {Zhou}, {Lou}, {Wu}, {Tang}, \&
  {He}}]{li2014}
{Li}, D.~L., {Esimbek}, J., {Zhou}, J.~J., {et~al.} 2014, \aap, 567, A10

\bibitem[{{Li} {et~al.}(2015){Li}, {Yuen}, {Otto}, {Leung}, {Sridharan},
  {Zhang}, {Liu}, {Tang}, \& {Qiu}}]{li15}
{Li}, H.-B., {Yuen}, K.~H., {Otto}, F., {et~al.} 2015, \nat, 520, 518

\bibitem[{{Li} \& {Klein}(2019)}]{li_klein2019}
{Li}, P.~S., \& {Klein}, R.~I. 2019, \mnras, 485, 4509

\bibitem[{{Lindegren} {et~al.}(2018){Lindegren}, {Hern{\'a}ndez}, {Bombrun},
  {Klioner}, {Bastian}, {Ramos-Lerate}, {de Torres}, {Steidelm{\"u}ller},
  {Stephenson}, \& {Hobbs}}]{lindegren2018}
{Lindegren}, L., {Hern{\'a}ndez}, J., {Bombrun}, A., {et~al.} 2018, \aap, 616,
  A2

\bibitem[{{Lippok} {et~al.}(2013){Lippok}, {Launhardt}, {Semenov}, {Stutz},
  {Balog}, {Henning}, {Krause}, {Linz}, {Nielbock}, {Pavlyuchenkov},
  {Schmalzl}, {Schmiedeke}, \& {Bieging}}]{lippok13}
{Lippok}, N., {Launhardt}, R., {Semenov}, D., {et~al.} 2013, \aap, 560, A41

\bibitem[{{Liu} {et~al.}(2019){Liu}, {Stutz}, \& {Yuan}}]{liu19}
{Liu}, H.-L., {Stutz}, A., \& {Yuan}, J.-H. 2019, \mnras, 487, 1259

\bibitem[{{Liu} {et~al.}(2018){Liu}, {Li}, {Juvela}, {Kim}, {Evans}, {Di
  Francesco}, {Liu}, {Yuan}, {Tatematsu}, {Zhang}, {Ward-Thompson}, {Fuller},
  {Goldsmith}, {Koch}, {Sanhueza}, {Ristorcelli}, {Kang}, {Chen}, {Hirano},
  {Wu}, {Sokolov}, {Lee}, {White}, {Wang}, {Eden}, {Li}, {Thompson}, {Pattle},
  {Soam}, {Nasedkin}, {Kim}, {Kim}, {Lai}, {Park}, {Qiu}, {Zhang}, {Alina},
  {Eswaraiah}, {Falgarone}, {Fich}, {Greaves}, {Gu}, {Kwon}, {Li}, {Malinen},
  {Montier}, {Parsons}, {Qin}, {Rawlings}, {Ren}, {Tang}, {Tang}, {Toth},
  {Wang}, {Wouterloot}, {Yi}, \& {Zhang}}]{liu18}
{Liu}, T., {Li}, P.~S., {Juvela}, M., {et~al.} 2018, \apj, 859, 151

\bibitem[{{Lombardi} {et~al.}(2014){Lombardi}, {Bouy}, {Alves}, \&
  {Lada}}]{lombardi14}
{Lombardi}, M., {Bouy}, H., {Alves}, J., \& {Lada}, C.~J. 2014, \aap, 566, A45

\bibitem[{{Matthews} \& {Wilson}(2000)}]{matthews2000}
{Matthews}, B.~C., \& {Wilson}, C.~D. 2000, \apj, 531, 868

\bibitem[{{Megeath} {et~al.}(2012){Megeath}, {Gutermuth}, {Muzerolle},
  {Kryukova}, {Flaherty}, {Hora}, {Allen}, {Hartmann}, {Myers}, {Pipher},
  {Stauffer}, {Young}, \& {Fazio}}]{megeath12}
{Megeath}, S.~T., {Gutermuth}, R., {Muzerolle}, J., {et~al.} 2012, \aj, 144,
  192

\bibitem[{{Miville-Desch{\^e}nes} {et~al.}(2017){Miville-Desch{\^e}nes},
  {Murray}, \& {Lee}}]{miville-deschenes16}
{Miville-Desch{\^e}nes}, M.-A., {Murray}, N., \& {Lee}, E.~J. 2017, \apj, 834,
  57

\bibitem[{{Motte} {et~al.}(2018){Motte}, {Bontemps}, \& {Louvet}}]{motte2017}
{Motte}, F., {Bontemps}, S., \& {Louvet}, F. 2018, \araa, 56, 41

\bibitem[{{Myers} {et~al.}(2000){Myers}, {Evans}, \& {Ohashi}}]{myers2000}
{Myers}, P.~C., {Evans}, N.~J., I., \& {Ohashi}, N. 2000, in Protostars and
  Planets IV, ed. V.~{Mannings}, A.~P. {Boss}, \& S.~S. {Russell}, 217

\bibitem[{{Nguyen Luong} {et~al.}(2011){Nguyen Luong}, {Motte}, {Hennemann},
  {Hill}, {Rygl}, {Schneider}, {Bontemps}, {Men'shchikov}, {Andr{\'e}},
  {Peretto}, {Anderson}, {Arzoumanian}, {Deharveng}, {Didelon}, {di Francesco},
  {Griffin}, {Kirk}, {K{\"o}nyves}, {Martin}, {Maury}, {Minier}, {Molinari},
  {Pestalozzi}, {Pezzuto}, {Reid}, {Roussel}, {Sauvage}, {Schuller}, {Testi},
  {Ward-Thompson}, {White}, \& {Zavagno}}]{nguyen11}
{Nguyen Luong}, Q., {Motte}, F., {Hennemann}, M., {et~al.} 2011, \aap, 535, A76

\bibitem[{{Olmi} \& {Testi}(2002)}]{olmi02}
{Olmi}, L., \& {Testi}, L. 2002, \aap, 392, 1053

\bibitem[{{Ossenkopf} \& {Henning}(1994)}]{ossen94}
{Ossenkopf}, V., \& {Henning}, T. 1994, \aap, 291, 943

\bibitem[{{Penoyre} {et~al.}(2020){Penoyre}, {Belokurov}, {Wyn Evans},
  {Everall}, \& {Koposov}}]{penoyre20}
{Penoyre}, Z., {Belokurov}, V., {Wyn Evans}, N., {Everall}, A., \& {Koposov},
  S.~E. 2020, \mnras, 495, 321

\bibitem[{{Planck Collaboration} {et~al.}(2014){Planck Collaboration},
  {Abergel}, {Ade}, {Aghanim}, {Alves}, {Aniano}, {Armitage-Caplan}, {Arnaud},
  {Ashdown}, {Atrio-Barand ela}, \& et~al.}]{planck13}
{Planck Collaboration}, {Abergel}, A., {Ade}, P.~A.~R., {et~al.} 2014, \aap,
  571, A11

\bibitem[{{Planck Collaboration} {et~al.}(2016){Planck Collaboration}, {Ade},
  {Aghanim}, {Alves}, {Arnaud}, {Arzoumanian}, {Ashdown}, {Aumont},
  {Baccigalupi}, {Band ay}, {Barreiro}, {Bartolo}, {Battaner}, {Benabed},
  {Beno{\^\i}t}, {Benoit-L{\'e}vy}, {Bernard}, {Bersanelli}, {Bielewicz},
  {Bock}, {Bonavera}, {Bond}, {Borrill}, {Bouchet}, {Boulanger}, {Bracco},
  {Burigana}, {Calabrese}, {Cardoso}, {Catalano}, {Chiang}, {Christensen},
  {Colombo}, {Combet}, {Couchot}, {Crill}, {Curto}, {Cuttaia}, {Danese},
  {Davies}, {Davis}, {de Bernardis}, {de Rosa}, {de Zotti}, {Delabrouille},
  {Dickinson}, {Diego}, {Dole}, {Donzelli}, {Dor{\'e}}, {Douspis}, {Ducout},
  {Dupac}, {Efstathiou}, {Elsner}, {En{\ss}lin}, {Eriksen},
  {Falceta-Gon{\c{c}}alves}, {Falgarone}, {Ferri{\`e}re}, {Finelli}, {Forni},
  {Frailis}, {Fraisse}, {Franceschi}, {Frejsel}, {Galeotta}, {Galli}, {Ganga},
  {Ghosh}, {Giard}, {Gjerl{\o}w}, {Gonz{\'a}lez-Nuevo}, {G{\'o}rski},
  {Gregorio}, {Gruppuso}, {Gudmundsson}, {Guillet}, {Harrison}, {Helou},
  {Hennebelle}, {Henrot-Versill{\'e}}, {Hern{\'a}ndez-Monteagudo}, {Herranz},
  {Hildebrand t}, {Hivon}, {Holmes}, {Hornstrup}, {Huffenberger}, {Hurier},
  {Jaffe}, {Jaffe}, {Jones}, {Juvela}, {Keih{\"a}nen}, {Keskitalo}, {Kisner},
  {Knoche}, {Kunz}, {Kurki-Suonio}, {Lagache}, {Lamarre}, {Lasenby},
  {Lattanzi}, {Lawrence}, {Leonardi}, {Levrier}, {Liguori}, {Lilje},
  {Linden-V{\o}rnle}, {L{\'o}pez-Caniego}, {Lubin}, {Mac{\'\i}as-P{\'e}rez},
  {Maino}, {Mandolesi}, {Mangilli}, {Maris}, {Martin},
  {Mart{\'\i}nez-Gonz{\'a}lez}, {Masi}, {Matarrese}, {Melchiorri}, {Mendes},
  {Mennella}, {Migliaccio}, {Miville-Desch{\^e}nes}, {Moneti}, {Montier},
  {Morgante}, {Mortlock}, {Munshi}, {Murphy}, {Naselsky}, {Nati},
  {Netterfield}, {Noviello}, {Novikov}, {Novikov}, {Oppermann}, {Oxborrow},
  {Pagano}, {Pajot}, {Paladini}, {Paoletti}, {Pasian}, {Perotto}, {Pettorino},
  {Piacentini}, {Piat}, {Pierpaoli}, {Pietrobon}, {Plaszczynski},
  {Pointecouteau}, {Polenta}, {Ponthieu}, {Pratt}, {Prunet}, {Puget}, {Rachen},
  {Reinecke}, {Remazeilles}, {Renault}, {Renzi}, {Ristorcelli}, {Rocha},
  {Rossetti}, {Roudier}, {Rubi{\~n}o-Mart{\'\i}n}, {Rusholme}, {Sandri},
  {Santos}, {Savelainen}, {Savini}, {Scott}, {Soler}, {Stolyarov}, {Sudiwala},
  {Sutton}, {Suur-Uski}, {Sygnet}, {Tauber}, {Terenzi}, {Toffolatti}, {Tomasi},
  {Tristram}, {Tucci}, {Umana}, {Valenziano}, {Valiviita}, {Van Tent},
  {Vielva}, {Villa}, {Wade}, {Wandelt}, {Wehus}, {Ysard}, {Yvon}, \&
  {Zonca}}]{planck16}
{Planck Collaboration}, {Ade}, P.~A.~R., {Aghanim}, N., {et~al.} 2016, \aap,
  586, A138

\bibitem[{{Poglitsch} {et~al.}(2010){Poglitsch}, {Waelkens}, {Geis},
  {Feuchtgruber}, {Vandenbussche}, {Rodriguez}, {Krause}, {Renotte}, {van
  Hoof}, {Saraceno}, {Cepa}, {Kerschbaum}, {Agn{\`e}se}, {Ali}, {Altieri},
  {Andreani}, {Augueres}, {Balog}, {Barl}, {Bauer}, {Belbachir}, {Benedettini},
  {Billot}, {Boulade}, {Bischof}, {Blommaert}, {Callut}, {Cara}, {Cerulli},
  {Cesarsky}, {Contursi}, {Creten}, {De Meester}, {Doublier}, {Doumayrou},
  {Duband}, {Exter}, {Genzel}, {Gillis}, {Gr{\"o}zinger}, {Henning},
  {Herreros}, {Huygen}, {Inguscio}, {Jakob}, {Jamar}, {Jean}, {de Jong},
  {Katterloher}, {Kiss}, {Klaas}, {Lemke}, {Lutz}, {Madden}, {Marquet},
  {Martignac}, {Mazy}, {Merken}, {Montfort}, {Morbidelli}, {M{\"u}ller},
  {Nielbock}, {Okumura}, {Orfei}, {Ottensamer}, {Pezzuto}, {Popesso},
  {Putzeys}, {Regibo}, {Reveret}, {Royer}, {Sauvage}, {Schreiber}, {Stegmaier},
  {Schmitt}, {Schubert}, {Sturm}, {Thiel}, {Tofani}, {Vavrek}, {Wetzstein},
  {Wieprecht}, \& {Wiezorrek}}]{pog10}
{Poglitsch}, A., {Waelkens}, C., {Geis}, N., {et~al.} 2010, \aap, 518, L2

\bibitem[{{Priestley} \& {Whitworth}(2020)}]{priestly2020}
{Priestley}, F.~D., \& {Whitworth}, A.~P. 2020, \mnras, 499, 3728

\bibitem[{{Rao} {et~al.}(2020){Rao}, {Gandhi}, {Knigge}, {Paice}, {Leigh}, \&
  {Boubert}}]{rao20}
{Rao}, A., {Gandhi}, P., {Knigge}, C., {et~al.} 2020, \mnras, 495, 1491

\bibitem[{{Reissl} {et~al.}(2018{\natexlab{a}}){Reissl}, {Stutz}, {Brauer},
  {Pellegrini}, {Schleicher}, \& {Klessen}}]{reissl18a}
{Reissl}, S., {Stutz}, A.~M., {Brauer}, R., {et~al.} 2018{\natexlab{a}},
  \mnras, 481, 2507

\bibitem[{{Reissl} {et~al.}(2020){Reissl}, {Stutz}, {Klessen}, {Seifried}, \&
  {Walch}}]{reissl20}
{Reissl}, S., {Stutz}, A.~M., {Klessen}, R.~S., {Seifried}, D., \& {Walch}, S.
  2020, arXiv e-prints, arXiv:2009.04201

\bibitem[{{Reissl} {et~al.}(2018{\natexlab{b}}){Reissl}, {Wolf}, \&
  {Brauer}}]{reissl18b}
{Reissl}, S., {Wolf}, S., \& {Brauer}, R. 2018{\natexlab{b}}, {POLARIS:
  POLArized RadIation Simulator}, ascl:1807.001

\bibitem[{{Schleicher} \& {Stutz}(2018)}]{schleicher2018}
{Schleicher}, D. R.~G., \& {Stutz}, A. 2018, \mnras, 475, 121

\bibitem[{{Seifried} {et~al.}(2020){Seifried}, {Haid}, {Walch}, {Borchert}, \&
  {Bisbas}}]{seifried20}
{Seifried}, D., {Haid}, S., {Walch}, S., {Borchert}, E.~M.~A., \& {Bisbas},
  T.~G. 2020, \mnras, 492, 1465

\bibitem[{{Shimajiri} {et~al.}(2019){Shimajiri}, {Andr{\'e}}, {Palmeirim},
  {Arzoumanian}, {Bracco}, {K{\"o}nyves}, {Ntormousi}, \&
  {Ladjelate}}]{shimajiri19}
{Shimajiri}, Y., {Andr{\'e}}, P., {Palmeirim}, P., {et~al.} 2019, \aap, 623,
  A16

\bibitem[{{Simon} {et~al.}(2006){Simon}, {Rathborne}, {Shah}, {Jackson}, \&
  {Chambers}}]{simon06}
{Simon}, R., {Rathborne}, J.~M., {Shah}, R.~Y., {Jackson}, J.~M., \&
  {Chambers}, E.~T. 2006, \apj, 653, 1325

\bibitem[{{Smith} {et~al.}(2012){Smith}, {Shetty}, {Stutz}, \&
  {Klessen}}]{smith12}
{Smith}, R.~J., {Shetty}, R., {Stutz}, A.~M., \& {Klessen}, R.~S. 2012, \apj,
  750, 64

\bibitem[{{Sodroski} {et~al.}(1997){Sodroski}, {Odegard}, {Arendt}, {Dwek},
  {Weiland}, {Hauser}, \& {Kelsall}}]{sodroski97}
{Sodroski}, T.~J., {Odegard}, N., {Arendt}, R.~G., {et~al.} 1997, \apj, 480,
  173

\bibitem[{{Soler}(2019)}]{soler19}
{Soler}, J.~D. 2019, \aap, 629, A96

\bibitem[{{Springel} {et~al.}(2018){Springel}, {Pakmor}, {Pillepich},
  {Weinberger}, {Nelson}, {Hernquist}, {Vogelsberger}, {Genel}, {Torrey},
  {Marinacci}, \& {Naiman}}]{springel18}
{Springel}, V., {Pakmor}, R., {Pillepich}, A., {et~al.} 2018, \mnras, 475, 676

\bibitem[{{Stephens} {et~al.}(2017){Stephens}, {Dunham}, {Myers}, {Pokhrel},
  {Sadavoy}, {Vorobyov}, {Tobin}, {Pineda}, {Offner}, {Lee}, {Kristensen},
  {J{\o}rgensen}, {Goodman}, {Bourke}, {Arce}, \& {Plunkett}}]{stephens17}
{Stephens}, I.~W., {Dunham}, M.~M., {Myers}, P.~C., {et~al.} 2017, \apj, 846,
  16

\bibitem[{{Stutz} {et~al.}(2010){Stutz}, {Launhardt}, {Linz}, {Krause},
  {Henning}, {Kainulainen}, {Nielbock}, {Steinacker}, \& {Andr{\'e}}}]{stutz10}
{Stutz}, A., {Launhardt}, R., {Linz}, H., {et~al.} 2010, \aap, 518, L87

\bibitem[{{Stutz}(2018)}]{stutz18}
{Stutz}, A.~M. 2018, \mnras, 473, 4890

\bibitem[{{Stutz} {et~al.}(2018){Stutz}, {Gonzalez-Lobos}, \&
  {Gould}}]{stutz18b}
{Stutz}, A.~M., {Gonzalez-Lobos}, V.~I., \& {Gould}, A. 2018, arXiv e-prints,
  arXiv:1807.11496

\bibitem[{{Stutz} \& {Gould}(2016)}]{stutz16}
{Stutz}, A.~M., \& {Gould}, A. 2016, \aap, 590, A2

\bibitem[{{Stutz} \& {Kainulainen}(2015)}]{stutz15}
{Stutz}, A.~M., \& {Kainulainen}, J. 2015, \aap, 577, L6

\bibitem[{{Stutz} {et~al.}(2013){Stutz}, {Tobin}, {Stanke}, {Megeath},
  {Fischer}, {Robitaille}, {Henning}, {Ali}, {di Francesco}, {Furlan},
  {Hartmann}, {Osorio}, {Wilson}, {Allen}, {Krause}, \& {Manoj}}]{stutz13}
{Stutz}, A.~M., {Tobin}, J.~J., {Stanke}, T., {et~al.} 2013, \apj, 767, 36

\bibitem[{{Tafalla} {et~al.}(2002){Tafalla}, {Myers}, {Caselli}, {Walmsley}, \&
  {Comito}}]{tafalla2002}
{Tafalla}, M., {Myers}, P.~C., {Caselli}, P., {Walmsley}, C.~M., \& {Comito},
  C. 2002, \apj, 569, 815

\bibitem[{{Tahani} {et~al.}(2018){Tahani}, {Plume}, {Brown}, \&
  {Kainulainen}}]{tahani2018}
{Tahani}, M., {Plume}, R., {Brown}, J.~C., \& {Kainulainen}, J. 2018, \aap,
  614, A100

\bibitem[{{Tahani} {et~al.}(2019){Tahani}, {Plume}, {Brown}, {Soler}, \&
  {Kainulainen}}]{tahani2019}
{Tahani}, M., {Plume}, R., {Brown}, J.~C., {Soler}, J.~D., \& {Kainulainen}, J.
  2019, \aap, 632, A68

\bibitem[{{Terry} {et~al.}(2019){Terry}, {Battaglia}, {Basu}, {Beringue},
  {Bertoldi}, {Chapman}, {Choi}, {Cothard}, {Chung}, {Erler}, {Fich},
  {Foreman}, {Gallardo}, {Gao}, {Graf}, {Haynes}, {Herter}, {Hilton},
  {Hubmayr}, {Johnstone}, {Komatsu}, {Magnelli}, {Mauskopf}, {McMahon},
  {Meerburg}, {Meyers}, {Mittal}, {Niemack}, {Nikola}, {Parshley}, {Riechers},
  {Stacey}, {Stutzki}, {Vavagiakis}, {Viero}, \& {Vissers}}]{ccat}
{Terry}, H., {Battaglia}, N., {Basu}, K., {et~al.} 2019, in Bulletin of the
  American Astronomical Society, Vol.~51, 213

\bibitem[{{Uchida} {et~al.}(1991){Uchida}, {Fukui}, {Minoshima}, {Mizuno}, \&
  {Iwata}}]{uchida91}
{Uchida}, Y., {Fukui}, Y., {Minoshima}, Y., {Mizuno}, A., \& {Iwata}, T. 1991,
  \nat, 349, 140

\bibitem[{{Ungerechts} \& {Thaddeus}(1987)}]{ungerechts87}
{Ungerechts}, H., \& {Thaddeus}, P. 1987, \apjs, 63, 645

\bibitem[{{Wareing} {et~al.}(2020){Wareing}, {Pittard}, \& {Falle}}]{wareing20}
{Wareing}, C.~J., {Pittard}, J.~M., \& {Falle}, S.~A.~E.~G. 2020, \mnras,
  arXiv:2011.01321

\bibitem[{{Zinn} {et~al.}(2019){Zinn}, {Pinsonneault}, {Huber}, \&
  {Stello}}]{zinn19}
{Zinn}, J.~C., {Pinsonneault}, M.~H., {Huber}, D., \& {Stello}, D. 2019, \apj,
  878, 136

\bibitem[{{Zucker} {et~al.}(2019){Zucker}, {Speagle}, {Schlafly}, {Green},
  {Finkbeiner}, {Goodman}, \& {Alves}}]{zucker2019}
{Zucker}, C., {Speagle}, J.~S., {Schlafly}, E.~F., {et~al.} 2019, \apj, 879,
  125

\end{thebibliography}

\setcounter{figure}{0} 
\renewcommand{\thefigure}{A.\arabic{figure}}
\renewcommand{\theHfigure}{A.\arabic{figure}}

\begin{figure*}[!b]
    \begin{minipage}{\textwidth}
    \centering
    \includegraphics[width=\textwidth]{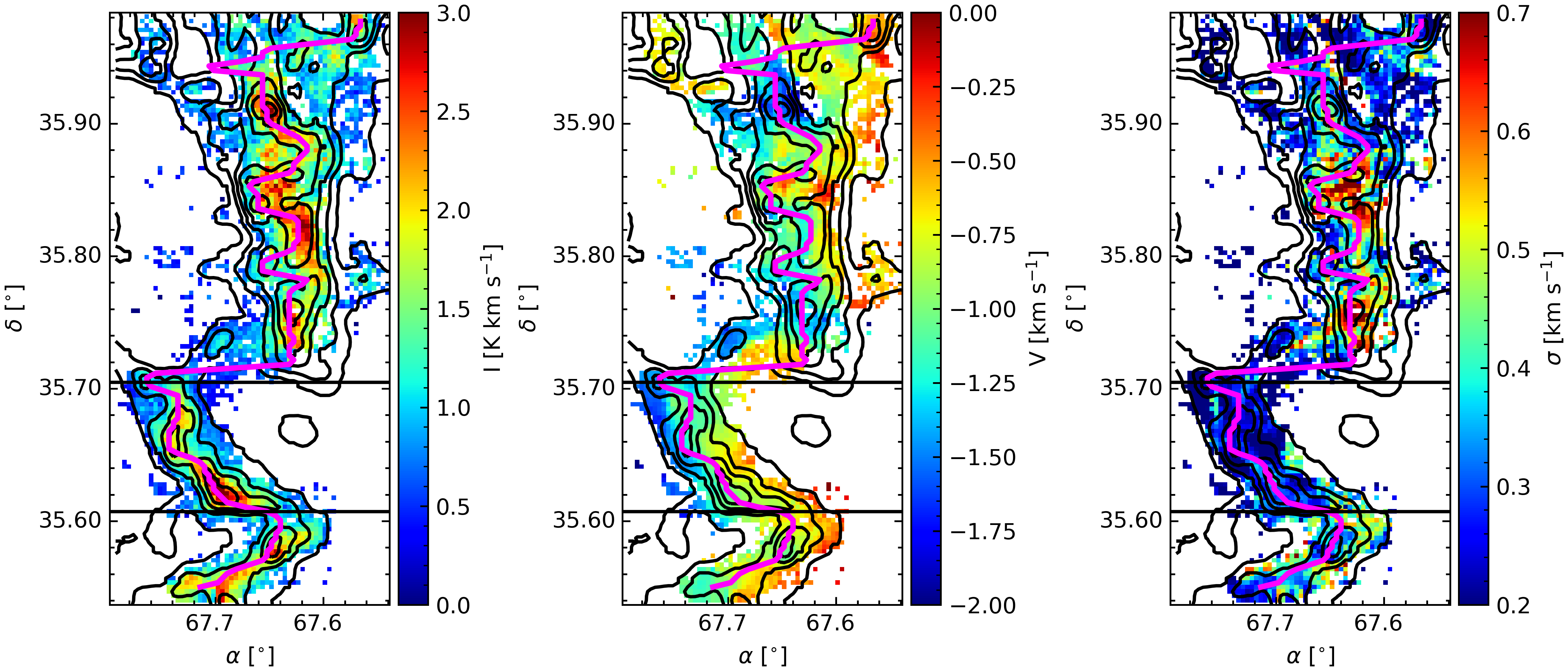}
    \includegraphics[width=\textwidth]{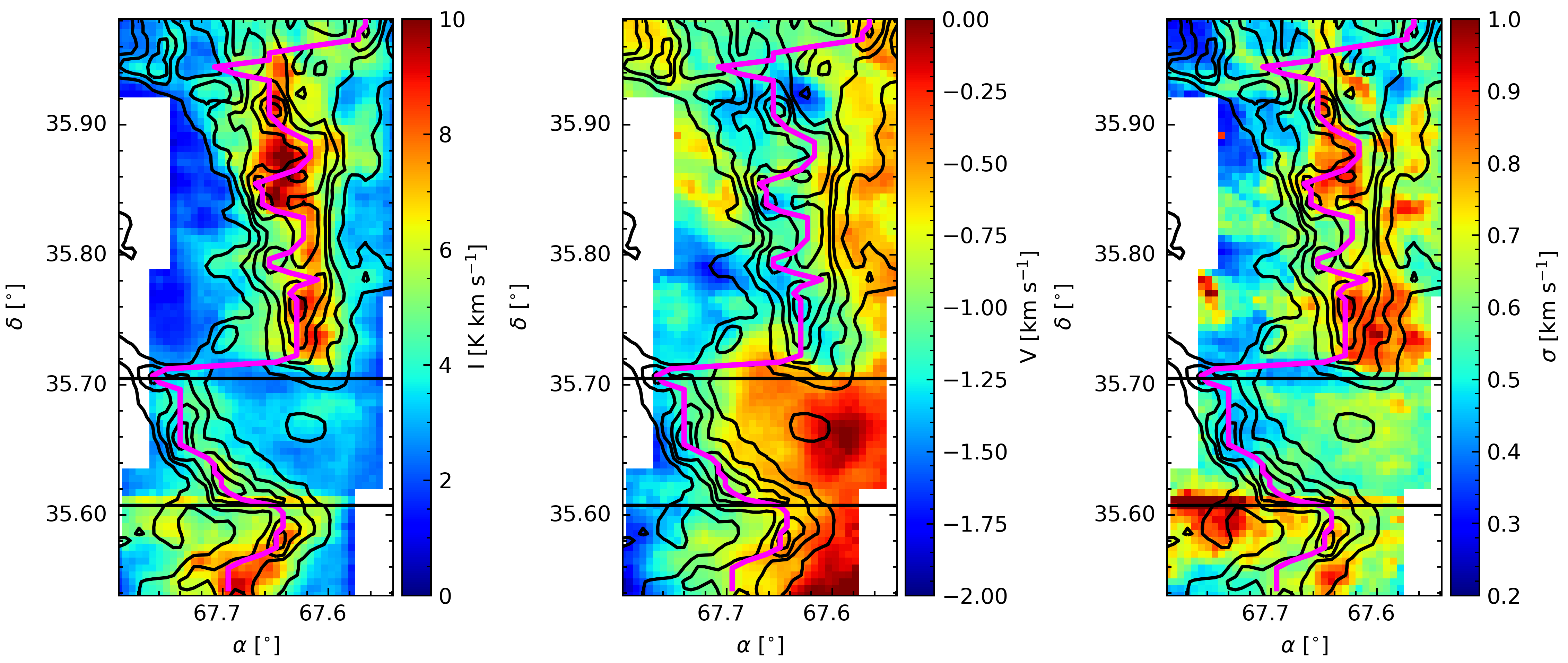}
    \caption{\co (top) and \sco (bottom) moments 0, 1, and 2 maps
      (left to right, respectively). The horizontal black lines
      indicate the extent of the South region (see text).  The magenta
      curve indicates the \nh ridgeline. The black contours follow
      $log_{10}(N_{\rm H}/\rm{cm}^{-2}) = 21.9, 22.1, 22.3, 22.4,
      22.6$, and $22.8$ values from the \herschel maps.}
    \label{fig:tracers_moments}
    \end{minipage}
\end{figure*}
\appendix

\section{Moment maps}
\label{appendix:a}

In Figure~\ref{fig:tracers_moments} we show the first 3 moment maps
(from left to right) for \co (top) and \sco (bottom). These figures
show that the filament is well traced in \co, but less so in \sco, a
fact that is particularly obvious in the velocity and line width
(moments 1 and 2) maps.

We find a larger scale \sco gradient in the same region we are
  analyzing.  It is tempting to associate the \sco gradient with the
  \co gradient, but as the moment maps show, the \sco tracer is
  simply not isolating the filament structure (including the radial
  velocities) nearly as well as the \co. For example, the filament is
  barely detected in the \sco moments 1 and 2 maps, and is only
  marginally detected in the moment 0 map; moreover the \sco gradient
  (not shown) is of a different magnitude as the \co, likely because
  the \sco is simply not tracing the filament but instead the broader
  and less well defined cloud environment.  Moreover, the rotational
  signature seen in Figure~\ref{fig:slopes} is not observed in the 
  immediately adjacent North region, as can be appreciated from the 
  moment maps. This is consistent with our interpretation that in 
  the North gravity has already take over, as evidenced by the 
  protostar content.  In summary, other lower density and optically 
  thick tracers are simply inadequate for isolating the filament 
  and its velocities.
  
\FloatBarrier

\section{Temperature profile}
\label{appendix:b}

In Figure~\ref{fig:fit_temp} we show the average projected radial
temperature profile (black solid line) of CMC/L1482.  We calculate 
this profile following the ridgeline (see above).  We also present 
the east and west temperature profiles.  Inside the radial range 
where the profile is used (see \S~\ref{sub:mach}), that is at 
r\,$\lesssim 1$\,pc, the profile is approximately symmetric and 
variations are small. Therefore we fit the average temperature 
profile with a softened power-law (red solid line).  We obtain 
the following best-fit temperature profile:

\begin{eqnarray}
\rm{T}(r) & = & 12.3 \ (1+(r/a)^2)^{0.04}\,\rm{K};
\qquad \textit{a} = 0.9~\pc.
\label{eq:temp}
\end{eqnarray}

\setcounter{figure}{0} 
\renewcommand{\thefigure}{B.\arabic{figure}}
\renewcommand{\theHfigure}{B\arabic{figure}}

\begin{figure}[h]
    \begin{minipage}{\columnwidth}
   \vspace{1cm}
    \centering
    \includegraphics[width=\columnwidth]{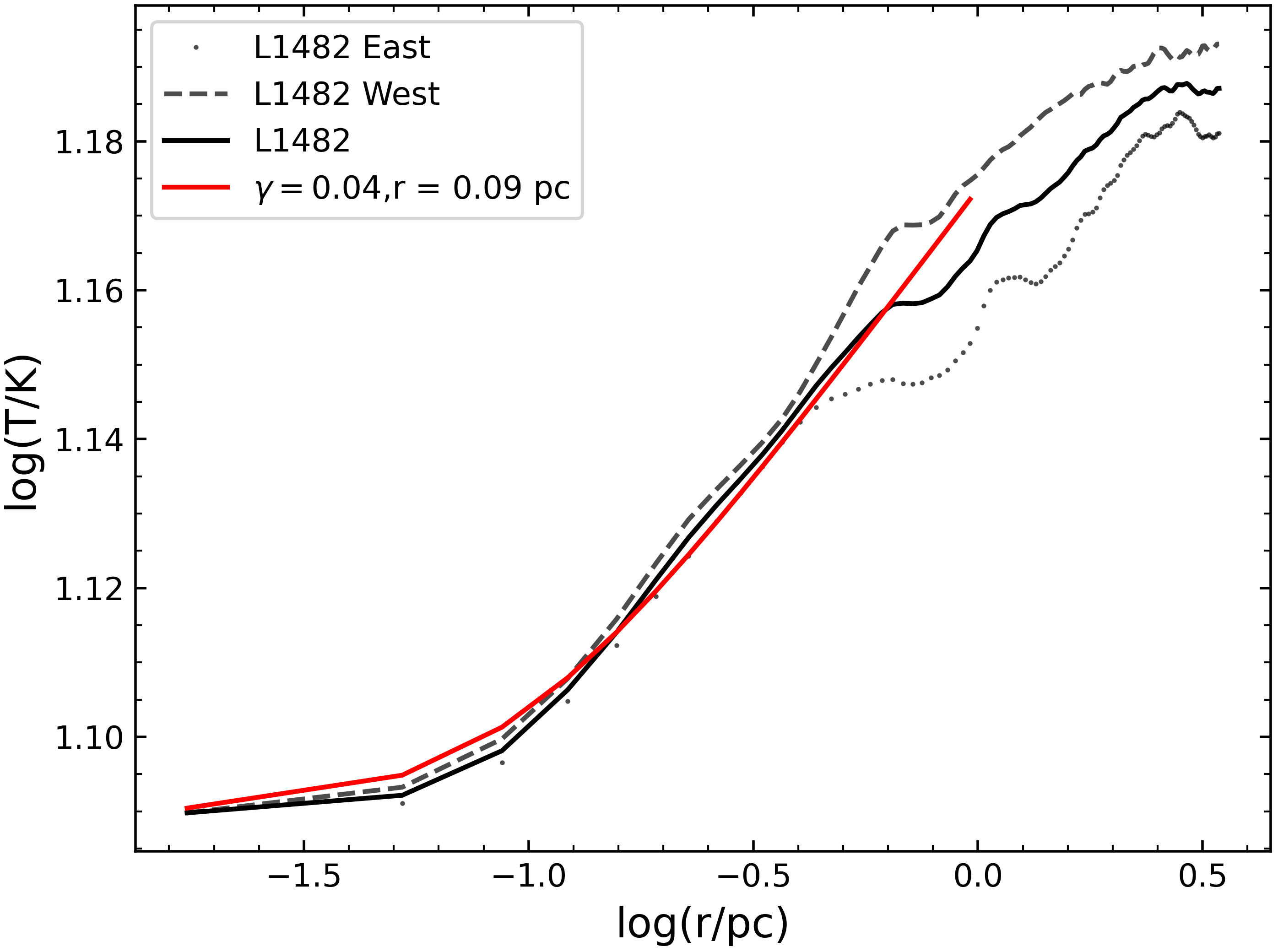}
    \caption{Average CMC/L1482 temperature profile derived from the
      \herschel temperature map (black solid line).  Dotted (dashed)
      curve corresponds to the east (west) profile.  The profiles are
      highly symmetric over the inner width of the filament.  The red
      curve shows the best-fit (see Eq.~\ref{eq:temp}) to the solid black
      line, and is plotted over the radial range that it is used in
      \S\ref{sub:mach} and in
      Figure~\ref{fig:mach_comparison}.}
    \label{fig:fit_temp}
    \end{minipage}
\end{figure}

\end{document}